\documentclass[12pt]{article}
\pdfoutput=1
\usepackage{amsmath,amssymb,amsthm,amsxtra,overpic,bbm,bm,epsfig,subfigure}
\usepackage{hyperref}
\usepackage{graphicx}
\usepackage{color}
\usepackage{array}
\usepackage{comment}
\usepackage{epstopdf}
\usepackage{float}
\usepackage{multirow}
\numberwithin{equation}{section}
\usepackage{cite}
\usepackage{hyperref}
\usepackage{slashed,stmaryrd}
\usepackage{bbm}
\usepackage{lscape}%
\usepackage{array}
\usepackage{booktabs}%

\def\thefootnote{\fnsymbol{footnote}}

\begin{document}

\vspace{0.2cm}

\begin{center}
{\Large\bf Muon $(g-2)$ and Flavor Puzzles in the $U(1)^{}_{X}$-gauged Leptoquark Model}
\end{center}

\vspace{0.2cm}

\begin{center}
	{\bf Xin Wang}~$^{a,~b}$~\footnote{E-mail: wangx@ihep.ac.cn}
	\\
	\vspace{0.2cm}
	{\small
		$^a$Institute of High Energy Physics, Chinese Academy of Sciences, Beijing 100049, China\\
		$^b$School of Physical Sciences, University of Chinese Academy of Sciences, Beijing 100049, China}
\end{center}

\vspace{1.5cm}

\begin{abstract}
We present an economical model where an $S^{}_1$ leptoquark and an anomaly-free $U(1)^{}_X$ gauge symmetry with $X = B^{}_3-2L^{}_\mu/3-L^{}_\tau/3$ are introduced, to account for the muon anomalous magnetic moment $a^{}_\mu \equiv (g^{}_\mu-2)$ and flavor puzzles including  $R^{}_{K^{(\ast)_{}}}$ and $R^{}_{D^{(\ast)_{}}}$ anomalies together with quark and lepton flavor mixing. The $Z^\prime_{}$ gauge boson associated with the $U(1)^{}_X$ symmetry is responsible for the $R^{}_{K^{(\ast)_{}}}$ anomaly. Meanwhile, the specific flavor mixing patterns of quarks and leptons can be generated after the spontaneous breakdown of the $U(1)^{}_X$ gauge symmetry via the Froggatt-Nielsen mechanism. The $S^{}_1$ leptoquark which is also charged under the $U(1)^{}_X$ gauge symmetry can simultaneously explain the latest muon $(g-2)$ result and the $R^{}_{D^{(\ast)_{}}}$ anomaly. In addition, we also discuss several other experimental constraints on our model.

\end{abstract}

\def\thefootnote{\arabic{footnote}}
\setcounter{footnote}{0}

\newpage

\section{Introduction}
Although a great number of experimental observations have proved that the Standard Model (SM) of particle physics can give an elegant description of the interactions among elementary particles in nature, the discovery of neutrino oscillations provides the firm evidence of tiny neutrino masses and significant lepton flavor mixing which are completely unexplained in the SM, and thus is a clear signal for new physics (NP) beyond the SM~\cite{Xing:2011zza,Xing:2019vks}.  On the other hand, hints of the deviations from the SM predictions could also be found in several precise measurements of the fundamental properties of the SM particles. 

Recently, the Fermi National Accelerator Laboratory Muon $(g-2)$ experiment reported the precise measurement of the muon anomalous magnetic moment $a^{}_\mu \equiv (g^{}_\mu-2)/2$ 
\begin{eqnarray}
\Delta a^{}_{\mu}=a_{\mu}^{\rm exp}-a_{\mu}^{\rm SM}=(251 \pm 59) \times 10^{-11}_{} \; ,
\label{eq:amuexp}
\end{eqnarray}
which combined with the final result from the E821 experiment at Brookhaven National Laboratory~\cite{Bennett:2006fi} shows a $4.2\sigma$ discrepancy~\cite{Abi:2021gix} with the SM theoretical prediction~\cite{Aoyama:2020ynm} (based on Refs.~\cite{Aoyama:2012wk,Aoyama:2019ryr,Czarnecki:2002nt,Gnendiger:2013pva,Davier:2017zfy,Keshavarzi:2018mgv,Colangelo:2018mtw,Hoferichter:2019mqg,Davier:2019can,Keshavarzi:2019abf,Kurz:2014wya,Melnikov:2003xd,Masjuan:2017tvw,Colangelo:2017fiz,Hoferichter:2018kwz,Gerardin:2019vio,Bijnens:2019ghy,Colangelo:2019uex,Blum:2019ugy,Colangelo:2014qya}). Such a result, although needs to be further examined by more experimental data as well as more reliable theoretical calculations, has attracted a lot of attention~\cite{Crivellin:2021rbq,Athron:2021iuf,Wang:2021bcx,Yin:2021mls,VanBeekveld:2021tgn,Abdughani:2021pdc,Endo:2021zal,Baum:2021qzx,Han:2021ify,Cox:2021gqq,Bai:2021bau,Gu:2021mjd,Buen-Abad:2021fwq,Aboubrahim:2021rwz,Arcadi:2021cwg,Wang:2021fkn,Ge:2021cjz,Zhu:2021vlz,Han:2021gfu,Cadeddu:2021dqx,Ibe:2021cvf,Ferreira:2021gke,Calibbi:2021qto,Brdar:2021pla,Escribano:2021css,Zu:2021odn,Das:2021zea,Baer:2021aax,Babu:2021jnu,Buras:2021btx,Marzocca:2021azj,Du:2021zkq,Chiang:2021pma,Cen:2021iwv,Li:2021lnz,Ban:2021tos,Ma:2021fre,Alvarado:2021nxy,Dasgupta:2021dnl,Dutta:2021afo,Chang:2021axw,Zhou:2021vnf,Aboubrahim:2021phn,Zhang:2021dgl,Zheng:2021wnu,Arkani-Hamed:2021xlp,Dey:2021pyn,Cirigliano:2021peb,Eung:2021bef,Singirala:2021gok,Yu:2021suw,Borah:2021mri,Chowdhury:2021tnm,Cao:2021tuh,Allwicher:2021rtd}.

Independently, lepton flavor universality (LFU) violation may exist in some semi-leptonic rare decays of $B$-mesons. For the neutral-current $b \rightarrow s\ell \ell$ process, the observables of which we are most interested in are the ratios of branching fractions~\cite{Aaij:2017vbb,Aaij:2019wad,Abdesselam:2019wac,Abdesselam:2019lab}
\begin{eqnarray}
R^{}_{K^{(*)}_{}} \equiv \frac{{\cal B}(B \rightarrow K^{(\ast)}_{}\mu^+_{}\mu^-){}}{{\cal B}(B \rightarrow K^{(\ast)}_{}e^+_{}e^{-}_{})} \; ,
\label{eq:RKdef}
\end{eqnarray}
which are predicted to be one in the SM with high accuracy~\cite{Bordone:2016gaq}. The latest results for $R^{}_K$ released by the Large Hadron Collider beauty (LHCb) collaboration show that $R^{}_K = 0.846^{+0.042}_{-0.039}$ for $1.1~{\rm GeV}^2_{} < q^2_{} < 6~{\rm GeV}^2_{}$ with $q^2_{}$ being the di-muon invariant mass squared~\cite{Aaij:2021vac}, which confirms the previous observed discrepancy and increases the statistical significance. If we include the experimental observations of the branching fraction $R^{}_{K^\ast_{}}$ and the absolute branching ratio of the purely leptonic decay $B^0_s \rightarrow \mu^{+}_{} \mu^{-}_{}$, the combined significance of deviation can arrive at $4.7\sigma$ for precise NP hypothesis~\cite{Geng:2021nhg,Altmannshofer:2021qrr}. Furthermore, the previous results for $R^{}_{K^{\ast}_{}}$ from the LHCb collaboration are in tension with the SM prediction at the $10\%$ level~\cite{Aaij:2017vbb}. Apart from the $b \rightarrow s\ell\ell$ transition, LFU violation is also actively searched for in the charged-current $b \rightarrow c\tau\nu$ process. There exist long-standing deviations in the ratios
\begin{eqnarray}
R^{}_{D^{(*)}_{}} \equiv \frac{{\cal B}(B \rightarrow D^{(\ast)}_{}\tau \nu)}{{\cal B}(B \rightarrow D^{(\ast)}_{}\ell \nu)} \quad ({\rm with}~\ell = e, \mu) \; .
\label{eq:RDdef}
\end{eqnarray}
The combined averages of $R^{}_D$ and $R^{}_{D^\ast_{}}$ announced by the Babar~\cite{Lees:2012xj,Lees:2013uzd}, Belle~\cite{Huschle:2015rga,Hirose:2016wfn} and LHCb~\cite{LHCb:2017smo,LHCb:2017vlu} collaborations differ from the SM predictions~\cite{Bernlochner:2017jka} by $1.4\sigma$ and $2.7\sigma$, respectively~\cite{Amhis:2019ckw}.

The above flavor anomalies, if confirmed by more accurate measurements, will be likely to call for explanations from NP. One intriguing way is to introduce the flavor-dependent $U(1)^{}_X$ gauge symmetry~\cite{Baek:2001kca,Ma:2001md,Allanach:2015gkd, Crivellin:2018qmi,Crivellin:2015mga,Crivellin:2016ejn,Altmannshofer:2015mqa,Bonilla:2017lsq,Alonso:2017uky,Allanach:2020kss,Ko:2019tts,Borah:2020swo}. The $Z^\prime_{}$ gauge boson associated with the $U(1)^{}_X$ symmetry couples non-universally to different generations of quarks and leptons. Therefore, it induces flavor-changing neutral-current processes, which can be used to address either the muon $(g-2)$ or $R^{}_{K^{(\ast)}_{}}$ anomaly. Also, the spontaneous symmetry breaking of $U(1)^{}_X$ could lead to realistic fermion masses and flavor mixing~\cite{Branco:1988ex,Choubey:2004hn,Araki:2012ip,Asai:2017ryy,Kownacki:2016pmx}. However, the explanation of $R^{}_{K^{(\ast)}_{}}$ usually requires a heavy $Z^\prime_{}$ boson, while the mass of $Z^\prime_{}$ which gives the correct contribution to $\Delta a^{}_\mu$ can not be larger than $400~{\rm MeV}$ due to the constraints from the neutrino trident experiment~\cite{Altmannshofer:2014pba}. So these two anomalies cannot be reconciled in the same parameter space if only a single $U(1)^{}_X$ gauge symmetry is involved.

An alternative avenue is the leptoquark model. Leptoquarks are hypothetical particles that simultaneously couple to quarks and leptons, and can provide dynamical solutions to the muon anomalous magnetic moment as well as anomalies in semi-leptonic $B$-meson decays~\cite{Davidson:1993qk,Dorsner:2016wpm,Gherardi:2020qhc,Crivellin:2020mjs}. The muon anomalous magnetic moment and the $R^{}_{D^{(*)}_{}}$ anomaly can be explained by an $S^{}_1\,(\overline{\bf 3},{\bf 1},1/3)$ (with the numbers in parenthesis being the  $SU(3)\times SU(2)^{}_{\rm L}\times U(1)_{\rm Y}$ quantum numbers) or an  $R^{}_1\,(\overline{\bf 3},{\bf 2},7/6)$ leptoquark~\cite{Babu:2010vp,Saad:2020ucl,Popov:2019tyc,Dorsner:2020aaz,Crivellin:2020tsz,Sakaki:2013bfa,Iguro:2018vqb,Bigaran:2020jil}, and the $S^{}_3\,(\overline{\bf 3},{\bf 3},1/3)$ leptoquark is regarded as the candidate to solve the $R^{}_{K^{(*)}_{}}$ anomaly~\cite{Pas:2015hca,Cheung:2016fjo,Crivellin:2017zlb,Gripaios:2014tna,Bhattacharya:2016mcc,Dorsner:2017ufx}. Unfortunately, none of the scalar leptoquarks alone can accommodate all the anomalies mentioned above~\cite{Bauer:2015knc,Cai:2017wry,Popov:2016fzr,Angelescu:2018tyl,Bigaran:2019bqv,Dorsner:2019itg,Saad:2020ihm,Babu:2020hun,Becirevic:2016yqi}. Another drawback of leptoquark models is that the leptoquark in principle couples indiscriminately to all the generations of leptons, which may bring about lepton-flavor-violating (LFV) processes. To avoid this problem, the combinations of leptoquark models and $U(1)^{}_X$ gauge symmetries have been discussed in previous literature~\cite{Hambye:2017qix,Davighi:2020qqa,Greljo:2021xmg,Greljo:2021npi}. 

In this paper, we investigate the possibility to explain the muon $(g-2)$ anomaly and some flavor puzzles by adopting only one leptoquark in the $U(1)^{}_X$-gauged model. We propose a viable model where an $S^{}_1$ leptoquark and an anomaly-free $U(1)^{}_X$ gauge symmetry with $X = B^{}_3-2L^{}_\mu/3-L^{}_\tau/3$ ($B^{}_3$ and $L^{}_{\mu,\tau}$ are respectively the baryon number of the third-generation quarks, and the lepton numbers of the second- and third-generation leptons) are introduced. The $Z^\prime_{}$ boson associated with the $U(1)^{}_X$ symmetry can generate the sizable contribution to $R^{}_{K^{(*)}_{}}$ at the tree level, and the $S^{}_1$ leptoquark charged under the $U(1)^{}_X$ gauge symmetry is responsible for the explanations of the muon anomalous magnetic moment and the $R^{}_{D^{(*)}_{}}$ anomaly. The $U(1)^{}_X$ symmetry restricts the couplings between $S^{}_1$ and fermions, leaving only a small number of non-zero coupling constants. Taking also other experimental constraints into consideration, we perform a numerical analysis and find out the allowed parameter space of these coupling constants. Specific flavor mixing patterns of quarks and leptons can be generated via the Froggatt-Nielsen mechanism by introducing an SM singlet $\chi$~\cite{Froggatt:1978nt}. In particular, the effective neutrino mass matrix turns out to possess the two-zero texture ${\bf A}^{}_2$~\cite{Frampton:2002yf, Xing:2002ta,Xing:2002ap,Fritzsch:2011qv,Zhou:2015qua}, which gives rise to proper neutrino masses and mixing parameters in accord with neutrino oscillation experiments. 

The remaining part of this paper is organized as follows. In Sec.~\ref{sec:model} we construct the concrete $U(1)^{}_X$-gauged leptoquark model. The quark and lepton flavor mixing is studied in Sec.~\ref{sec:mix}. We illustrate how our model can account for muon $(g-2)$ and $B$-anomalies in Sec.~\ref{sec:sol}. In Sec.~\ref{sec:glo} we carry out a numerical analysis and obtain the allowed parameter space of our model. We summarize our main results in Sec.~\ref{sec:sum}. 

\section{The $U(1)^{}_{X}$-gauged leptoquark model}\label{sec:model}
It is generally possible to extend the SM gauge group $SU(3)^{}_{\rm c} \times SU(2)^{}_{\rm L} \times U(1)^{}_{\rm Y}$ by including an extra $U(1)^{}_X$ gauge symmetry, which should be non-anomalous if we require the whole theory to be self-consistent. In this paper, we consider the $U(1)^{}_X$ gauge symmetry with $X = B^{}_3-x L^{}_\mu-(1-x) L^{}_\tau$, which is anomaly-free in the extension of the SM with three right-handed neutrinos. As we shall show later, $x= 2/3$ is the only choice to simultaneously resolve the muon $(g-2)$ and $B$-anomalies in our model. The $Z^\prime_{}$ gauge boson associated with the $U(1)^{}_X$ gauge symmetry couples selectively to the third generation of quarks as well as the second and third generations of leptons, which can naturally address the $R^{}_{K^{(*)}_{}}$ anomaly. In order for the combined explanations of muon anomalous magnetic moment and the $R^{}_{D^{(*)}_{}}$ anomaly, an $S^{}_1$ leptoquark with the charge $Q^{}_X(S^{}_1)=1/3$ under the $U(1)^{}_X$ gauge symmetry is also introduced. Furthermore, it is evident that the flavor mixing patterns of quarks and leptons are restricted by the $U(1)^{}_X$ gauge symmetry. To be specific, the mixing matrix of quarks is described by a (1,2)-rotation while that of leptons is explicitly diagonal. On this account, we employ the Froggatt-Nielsen mechanism~\cite{Froggatt:1978nt} in both quark and lepton sectors. A flavon field $\chi$ which is neutral in the SM but with the charge $Q^{}_X(\chi) = 1/3$ under the $U(1)^{}_X$ symmetry is included to generate realistic quark and lepton flavor mixing. The charge assignments of the SM particles, three right-handed neutrinos $\nu^{}_{\alpha{\rm R}}$ (for $\alpha=e,\mu,\tau$), the $S^{}_1$ leptoquark and the flavon $\chi$ under $SU(3)^{}_{\rm c} \otimes SU(2)^{}_{\rm L} \otimes U(1)^{}_{\rm Y} \otimes U(1)^{}_X$ are listed in Table~\ref{table:charge}. Due to the gauge field $Z^\prime_{\mu}$, the modified covariant derivative can be written as
\begin{eqnarray}
D_{\mu}^{\prime} \equiv \partial^{}_{\mu}-{\rm i} g \tau^{a}_{} W_{\mu}^{a}-{\rm i} g^{\prime}_{} Y B^{}_{\mu}-{\rm i} g^{}_{Z^{\prime}_{}} Q_{X}^{} Z_{\mu}^{\prime} \; ,
\label{eq:coderivative}
\end{eqnarray}
where $Q^{}_X$ represents the charge under the $U(1)^{}_X$ symmetry and $g^{}_{Z^\prime_{}}$ is the new gauge coupling constant. It is straightforward to derive the neutral-current interactions related to the $Z^\prime_{}$ boson by implementing the covariant derivative in Eq.~(\ref{eq:coderivative}) on the SM fermions together with three right-handed neutrinos, namely,
\begin{eqnarray}
{\cal L}^{}_{\rm NC} &= & -\frac{g^{}_{Z^{\prime}_{}}}{3}\left(2\overline{\ell^{}_{\mu \mathrm{L}}} \gamma^{\mu}_{} \ell^{}_{\mu \mathrm{L}}+\overline{\ell^{}_{\tau \mathrm{L}}} \gamma^{\mu}_{} \ell^{}_{\tau \mathrm{L}}+2\overline{\mu^{}_{\mathrm{R}}} \gamma^{\mu}_{} \mu^{}_{\mathrm{R}}+\overline{\tau^{}_{\mathrm{R}}} \gamma^{\mu}_{} \tau^{}_{\mathrm{R}}+2\overline{\nu^{}_{\mu \mathrm{R}}} \gamma^{\mu}_{} \nu^{}_{\mu \mathrm{R}}+\overline{\nu^{}_{\tau \mathrm{R}}} \gamma^{\mu}_{} \nu^{}_{\tau \mathrm{R}}\right) Z_{\mu}^{\prime} \nonumber \\
&&+\frac{g^{}_{Z^{\prime}_{}}}{3}\left(\overline{Q^{}_{3{\rm L}}}\gamma^\mu_{}Q^{}_{3{\rm L}} + \overline{t^{}_{\rm R}}\gamma^\mu_{} t^{}_{\rm R}  +\overline{b^{}_{\rm R}}\gamma^\mu_{} b^{}_{\rm R}\right) Z^\prime_\mu \; ,
\label{eq:NC}
\end{eqnarray}
which is written in the flavor basis of quarks and leptons. It is not difficult to see that ${\cal L}^{}_{\rm NC}$ can induce the flavor-changing $b \rightarrow s$ transition if we transform it to the mass basis of quarks.

The $U(1)^{}_X$ and $SU(2)^{}_{\rm L} \otimes U(1)^{}_{\rm Y}$ symmetries can be spontaneously broken after $\chi$ and $H$ obtain their individual vacuum expectation values (vev's)
\begin{eqnarray}
\langle \chi \rangle = \frac{v^{}_\chi}{\sqrt{2}} \; , \quad    \langle H \rangle = \left( 0 \quad \dfrac{v^{}_H}{\sqrt{2}} \right)^{\rm T}_{}\; .
\label{eq:vev}
\end{eqnarray}
The SM Higgs doublet $H$ is neutral under the $U(1)^{}_X$ symmetry, whereas $\chi$ only possesses non-zero $U(1)^{}_X$ charge, so there seems no mass mixing between $Z$ and $Z^\prime_{}$. It should be noticed that the kinetic mixing term $\kappa Z^\prime_{\mu\nu} B^{\mu\nu}_{}$ (with $Z^\prime_{\mu\nu} \equiv \partial^{}_{\mu} Z^{\prime}_{\nu} - \partial^{}_{\nu} Z^{\prime}_{\mu}$ and $B^{}_{\mu\nu} \equiv \partial^{}_{\mu} B^{}_{\nu} - \partial^{}_{\nu} B^{}_{\mu}$ being defined) between the $U(1)^{}_{X}$ and $U(1)^{}_{\rm Y}$ gauge bosons could in general exist in the Lagrangian. Up to the first order of $\kappa$, the kinetic mixing angle is approximately proportional to $\kappa m^2_Z/(m^2_{Z^\prime_{}}-m^2_Z)$. As we will see later, the mass of $Z^\prime_{}$ in our model should be much heavier than the $Z$ gauge boson to explain the $R^{}_{K^{(\ast)}_{}}$ anomaly. This implies that the mixing angle between $Z$ and $Z^\prime_{}$ would be highly suppressed by their mass ratio, thus the kinetic mixing effect between $Z$ and $Z^\prime_{}$ is also negligibly small. In this regard, the mass of $Z$ remains unchanged while that of $Z^\prime_{}$ is simply given by $m^{}_{Z^\prime_{}} = g^{}_{Z^\prime_{}}v^{}_\chi/2$.

On the other hand, keeping the charge assignments in mind, we can write down the Lagrangian relevant for the masses of quarks and leptons
\begin{table}[t!]
	\centering
	\caption{Charge assignments of the SM particles, three right-handed neutrinos $\nu^{}_{\alpha{\rm R}}$ (for $\alpha=e,\mu,\tau$), the $S^{}_1$ leptoquark and the flavon $\chi$ under the  $SU(3)^{}_{\rm c} \otimes SU(2)^{}_{\rm L} \otimes U(1)^{}_{\rm Y} \otimes U(1)^{}_X$ gauge symmetry, where the indices $i=1,2$ of $\{Q^{}_{i{\rm L}},u^{}_{i{\rm R}},d^{}_{i{\rm R}}\}$ represent the first two generations of quarks, and $H$ denotes the SM Higgs doublet.}
	\vspace{0.3cm}
	\begin{tabular}{c|ccccccccc}
		\hline\hline
		& $Q^{}_{i{\rm L}}$ & $Q^{}_{3{\rm L}}$ & $u^{}_{i{\rm R}}$ & $d^{}_{i{\rm R}}$ & $t^{}_{\rm R}$ & $b^{}_{\rm R}$ & $\ell^{}_{e{\rm L}}$ & $\ell^{}_{\mu{\rm L}}$ & $\ell^{}_{\tau{\rm L}}$ \\
		\hline
		$SU(3)^{}_{\rm c}$ & {\bf 3} &{\bf 3}&{\bf 3}&{\bf 3}&{\bf 3}&{\bf 3}&{\bf 1}&{\bf 1}&{\bf 1}\\
		$SU(2)^{}_{\rm L}$ &{\bf 2}&{\bf 2}&{\bf 1}&{\bf 1}&{\bf 1}&{\bf 1}&{\bf 2}&{\bf 2}&{\bf 2}\\
		$U(1)^{}_{\rm Y}$ & $+1/6$ & $+1/6$ & $+2/3$ & $-1/3$ & $+2/3$ & $-1/3$ &  $-1/2$ & $-1/2$ & $-1/2$  \\
		$U(1)^{}_{X}$ & 0 & $+1/3$ & 0 & 0 & $+1/3$ & $+1/3$ & 0 & $-2/3$ &$-1/3$ \\
		\hline\hline
		& $e^{}_{\rm R}$ & $\mu^{}_{\rm R}$ & $\tau^{}_{\rm R}$ & $\nu^{}_{e {\rm R}}$ & $\nu^{}_{\mu {\rm R}}$ & $\nu^{}_{\tau {\rm R}}$ & $H$ & $\chi$ & $S^{}_1$\\
		\hline
		$SU(3)^{}_{\rm c}$ &{\bf 1}&{\bf 1}&{\bf 1}&{\bf 1}&{\bf 1}&{\bf 1}&{\bf 1}&{\bf 1}& $\overline{\bf 3}$\\
		$SU(2)^{}_{\rm L}$ &{\bf 2}&{\bf 2}&{\bf 2}&{\bf 1}&{\bf 1}&{\bf 1}&{\bf 2}&{\bf 1}& {\bf 1}\\
		$U(1)^{}_{\rm Y}$ &  $-1$ & $-1$ & $-1$ & $0$ & $0$ &$0$ & $+1/2$ & 0 & $+1/3$ \\
		$U(1)^{}_{X}$ &  0 & $-2/3$ & $-1/3$ & 0 & $-2/3$ & $-1/3$ & 0 & $+1/3$ & $+1/3$ \\
		\hline\hline
	\end{tabular}
	\label{table:charge}
	\vspace{0.3cm}
\end{table}
\begin{eqnarray}
	-{\cal L}^{}_{\rm Y} &=& y^{\rm u}_{ij}\overline{Q^{}_{i{\rm L}}}\widetilde{H}u^{}_{j{\rm R}} + y^{\rm u}_{33}\overline{Q^{}_{3{\rm L}}}\widetilde{H}t^{}_{\rm R} + y^{\rm d}_{ij}\overline{Q^{}_{i{\rm L}}} H d^{}_{j{\rm R}} + y^{\rm d}_{33}\overline{Q^{}_{3{\rm L}}} 
	H b^{}_{\rm R} \nonumber \\
	&& + y^l_e \overline{\ell^{}_{e {\rm L}}} H e^{}_{\rm R} +  y^l_\mu \overline{\ell^{}_{\mu {\rm L}}} H \mu^{}_{\rm R} + y^l_\tau \overline{\ell^{}_{\tau {\rm L}}} H \tau^{}_{\rm R} + y^\nu_e \overline{\ell^{}_{e {\rm L}}} \widetilde{H} \nu^{}_{e{\rm R}} +  y^\nu_\mu \overline{\ell^{}_{\mu {\rm L}}} \widetilde{H} \nu^{}_{\mu{\rm R}} + y^\nu_\tau \overline{\ell^{}_{\tau {\rm L}}} \widetilde{H} \nu^{}_{\tau{\rm R}} \nonumber \\
	&& + \widetilde{y}^{\rm u}_{3i} \frac{\chi}{\Lambda} \overline{Q^{}_{3{\rm L}}}\widetilde{H}u^{}_{j{\rm R}} + \widetilde{y}^{\rm u}_{i3} \frac{\chi^\dag_{}}{\Lambda} \overline{Q^{}_{i{\rm L}}}\widetilde{H}t^{}_{\rm R} + \widetilde{y}^{\rm d}_{3i} \frac{\chi}{\Lambda} \overline{Q^{}_{3{\rm L}}} H d^{}_{j{\rm R}} + \widetilde{y}^{\rm d}_{i3} \frac{\chi^\dag_{}}{\Lambda} \overline{Q^{}_{i{\rm L}}} H b^{}_{\rm R} + \frac{1}{2} m^{ee}_{\rm R}\overline{\nu^{\rm C}_{e{\rm R}}} \nu^{}_{e{\rm R}} \nonumber \\
	&& +\frac{\chi}{2} \left[y^{e\tau}_{\chi}\left(\overline{\nu^{\rm C}_{e{\rm R}}} \nu^{}_{\tau{\rm R}} + \overline{\nu^{\rm C}_{\tau{\rm R}}} \nu^{}_{e{\rm R}}\right) + y^{e\mu}_\chi \frac{\chi}{\Lambda}\left(\overline{\nu^{\rm C}_{e{\rm R}}} \nu^{}_{\mu{\rm R}} + \overline{\nu^{\rm C}_{\mu{\rm R}}} \nu^{}_{e{\rm R}}\right)+y^{\tau\tau}_\chi \frac{\chi}{\Lambda}\overline{\nu^{\rm C}_{\tau{\rm R}}} \nu^{}_{\tau{\rm R}}\right] + {\rm h.c.} \; ,
	\label{eq:Yuk}
\end{eqnarray}
where $\widetilde{H} \equiv {\rm i}\sigma^{}_2 H^\ast_{}$ and $\Lambda$ stands for the cut-off scale. It should be noticed that the off-diagonal Yukawa couplings in the lepton sector are set to be zero. This assumption is reasonable given that these couplings are suppressed by $\chi/\Lambda$ and thus could not lead to significant contributions to the lepton flavor mixing. It is then straightforward for us to derive the fermion mass matrices after the spontaneous breakdown of the $U(1)^{}_X$ and $SU(2)^{}_{\rm L} \otimes U(1)^{}_{\rm Y}$ gauge symmetries from Eq.~(\ref{eq:Yuk}). For the up- and down-type quark mass matrices, we have
\begin{eqnarray}
	M^{}_{\rm u} = \frac{v^{}_H}{\sqrt{2}}\left(
	\begin{matrix}
		y^{\rm u}_{11} & y^{\rm u}_{12} & \xi\widetilde{y}^{\rm u}_{13} \\
		y^{\rm u}_{21} & y^{\rm u}_{22} & \xi\widetilde{y}^{\rm u}_{23} \\
		\xi\widetilde{y}^{\rm u}_{31} & \xi\widetilde{y}^{\rm u}_{32} & y^{\rm u}_{33} \\
	\end{matrix}\right) \; , 
	\quad
	M^{}_{\rm d} = \frac{v^{}_H}{\sqrt{2}}\left(
	\begin{matrix}
		y^{\rm d}_{11} & y^{\rm d}_{12} & \xi\widetilde{y}^{\rm d}_{13} \\
		y^{\rm d}_{21} & y^{\rm d}_{22} & \xi\widetilde{y}^{\rm d}_{23} \\
		\xi\widetilde{y}^{\rm d}_{31} & \xi\widetilde{y}^{\rm d}_{32} & y^{\rm d}_{33} \\
	\end{matrix}\right) \; , 
	\label{eq:mass_quark}
\end{eqnarray}
where $\xi \equiv v^{}_\chi/\Lambda$ has been defined. Generally speaking, $v^{}_\chi$ should be smaller than the cut-off scale $\Lambda$, i.e., $\xi <1$. For illustration, in the rest part of this paper we assume $\xi = 0.1$. In the lepton sector, the charged-lepton mass matrix, the Dirac and Majorana neutrino mass matrices take the forms as
\begin{eqnarray}
M^{}_{l} = \frac{v^{}_H}{\sqrt{2}}\left(
\begin{matrix}
y^{l}_{e} & 0 & 0 \\
0 & y^l_{\mu} & 0 \\
0 & 0 & y^l_{\tau} \\
\end{matrix}\right) \; , 
\quad
M^{}_{\rm D} = \frac{v^{}_H}{\sqrt{2}}\left(
\begin{matrix}
y^{\nu}_{e} & 0 & 0 \\
0 & y^{\nu}_{\mu} & 0 \\
0 & 0 & y^{\nu}_{\tau} \\
\end{matrix}\right) \; , 
\quad 
M^{}_{\rm R} =\frac{v^{}_\chi}{\sqrt{2}} \left(
\begin{matrix}
\widehat{m}^{ee}_{\rm R} & y^{e\mu}_{\chi}\xi & y^{e\tau}_{\chi} \\
y^{e\mu}_{\chi}\xi & 0 & 0 \\
y^{e\tau}_{\chi} & 0 & y^{\tau\tau}_{\chi}\xi \\
\end{matrix}\right) \; ,
\label{eq:mass_lepton}
\end{eqnarray}
with $\widehat{m}^{ee}_{\rm R} \equiv m^{ee}_{\rm R}/v^{}_\chi$. Since $M^{}_l$ is diagonal, one can always make coupling constants $y^{l}_\alpha$ (for $\alpha = e, \mu, \tau$) real by redefining the phases of the right-handed charged-lepton fields. Similarly, the diagonal $M^{}_{\rm D}$ allows us to make $y^{\nu}_\alpha$ real by absorbing their phases into the left-handed lepton doublets. As for the Majorana neutrino mass matrix $M^{}_{\rm R}$, having the aid of the freedom of redefining the phases of right-handed neutrinos, we can remove the phases of $\widehat{m}^{ee}_{\rm R}$, $y^{e\mu}_{\chi}$ and one of $y^{e\tau}_{\chi}$ and $y^{\tau\tau}_{\chi}$. For definiteness, we choose $y^{e\tau}_{\chi}$ to be the only complex parameter in $M^{}_{\rm R}$, i.e., $y^{e\tau}_{\chi} = |y^{e\tau}_{\chi}| e^{{\rm i}\phi^{}_{e\tau}}$. With the help of the canonical seesaw formula $M^{}_\nu \approx -M^{}_{\rm D}M^{-1}_{\rm R}M^{\rm T}_{\rm D}$~\cite{Minkowski:1977sc, Yanagida:1979, GellMan1979, Mohapatra:1979ia}, we arrive at the effective neutrino mass matrix
\begin{eqnarray}
M^{}_\nu = \dfrac{v^2_H}{\sqrt{2}v^{}_\chi}\left(
\begin{matrix}
0 & \dfrac{y^\nu_e y^\nu_\mu}{y^{e\mu}_\chi\xi} & 0 \\
\dfrac{y^\nu_e y^\nu_\mu}{y^{e\mu}_\chi\xi} & \dfrac{-(y^\nu_{\mu})^2_{}[\widehat{m}^{ee}_{\rm R}-(y^{e\tau}_\chi)^2_{}]}{(y^{e\mu}_{\chi})^2_{}y^{\tau\tau}_\chi\xi^3_{}} & -\dfrac{y^{e\tau}_\chi y^\nu_\mu y^\nu_\tau}{y^{e\mu}_\chi y^{\tau\tau}_\chi \xi^2_{}} \\
0 & -\dfrac{y^{e\tau}_\chi y^\nu_\mu y^\nu_\tau}{y^{e\mu}_\chi y^{\tau\tau}_\chi \xi^2_{}} & \dfrac{(y^\nu_\tau)^2_{}}{y^{\tau\tau}_\chi \xi}
\end{matrix}\right) \; ,
\label{eq:mass_neutrino}
\end{eqnarray}
which is nothing but the ${\bf A}^{}_2$ two-zero texture. The flavor mixing patterns of quarks and leptons indicated by Eqs.~(\ref{eq:mass_quark}) and (\ref{eq:mass_neutrino}) will be studied in the next section.

Finally, the interactions between the $S^{}_1$ leptoquark and the SM fermions are given by
\begin{eqnarray}
	{\cal L}^{}_{\rm LQ} = \lambda^{\rm L}_{3\mu}\overline{Q^{\rm C}_{3{\rm L}}}\epsilon \ell^{}_{\mu {\rm L}} S^{}_1 + \lambda^{\rm L}_{2\tau}\overline{Q^{\rm C}_{2{\rm L}}}\epsilon \ell^{}_{\tau {\rm L}} S^{}_1 + \lambda^{\rm R}_{3\mu}\overline{t^{\rm C}_{\rm R}}\epsilon \mu^{}_{\rm R} S^{}_1 + \lambda^{\rm R}_{2\tau}\overline{u^{\rm C}_{\rm 2R}}\epsilon \tau^{}_{\rm R} S^{}_1 + {\rm h.c.} \; ,
	\label{eq:LQ}
\end{eqnarray}
where $\epsilon$ denotes the two-dimensional Levi-Civita symbol with $\epsilon^{}_{12} = -\epsilon^{}_{21} = 1$ and $\epsilon^{}_{11} = \epsilon^{}_{22} = 0$. Some remarks about Eq.~(\ref{eq:LQ}) are in order. First, as the $U(1)^{}_X$ symmetry tightly constrains the couplings among the $S^{}_1$ leptoquark, quarks and leptons, the number of non-zero coupling coefficients in the above equation is very limited. Second, the first two generations of quarks possess exactly the same quantum numbers under the $SU(3)^{}_{\rm c} \otimes SU(2)^{}_{\rm L} \otimes U(1)^{}_{\rm Y} \otimes U(1)^{}_X$ symmetry, so the couplings among $S^{}_1$, the first generation of quarks and third generation of leptons should also exist in Eq.~(\ref{eq:LQ}). Nevertheless, they only play a minor role in addressing the aforementioned anomalies and thus have been omitted. In fact, these couplings can be constrained by considering the processes like  $D^0_{}-\overline{D}{}^{0}_{}$ mixing. Third, the diquark coupling terms such as $\overline{Q^{\rm C}_{1{\rm L}}}\epsilon Q^{}_{3\rm L}S^\dag_1$ also comply with the SM and $U(1)^{}_X$ gauge symmetries. However, such terms may result in the dangerous proton decay, and can be forbidden by requiring the global baryon number conservation.

\section{Quark and lepton flavor mixing}\label{sec:mix}
In the most general case, each matrix in Eq.~(\ref{eq:mass_quark}) is an arbitrary $3\times3$ complex matrix which can be diagonalized by the unitary matrices $V^{{\rm L},{\rm R}}_{\rm u}$ or $V^{{\rm L},{\rm R}}_{\rm d}$ via $V^{{\rm L}\dag}_{\rm u} M^{}_{\rm u} V^{\rm R}_{\rm u} = {\rm Diag}\{m^{}_u,m^{}_c,m^{}_t\}$ or $V^{{\rm L}\dag}_{\rm d} M^{}_{\rm d} V^{\rm R}_{\rm d} = {\rm Diag}\{m^{}_d,m^{}_s,m^{}_b\}$. Then the Cabibbo-Kobayashi-Maskawa (CKM) matrix~\cite{Cabibbo:1963yz, Kobayashi:1973fv} is written as $V = V^{{\rm L}\dag}_{{\rm u}} V^{\rm L}_{\rm d}$. Nevertheless, since the elements $\xi\widetilde{y}^{\rm u,d}_{ij}$ are suppressed by the small parameter $\xi$, both $M^{}_{\rm u}$ and $M^{}_{\rm d}$ should be approximately diagonalized by the $(1,2)$-rotations. It naturally reminds us of the Fritzsch-Xing (FX) parametrization~\cite{Fritzsch:1997fw}
\begin{eqnarray}
V^{}_{\rm FX} = \left(
\begin{matrix}
c^{}_{\rm u} & s^{}_{\rm u} & 0 \\
-s^{}_{\rm u} & c^{}_{\rm u} & 0 \\
0 & 0 & 1 \\
\end{matrix}\right) \cdot
\left(
\begin{matrix}
e^{\rm -i\phi} & 0 & 0 \\
0 & c^{}_{\rm q} & s^{}_{\rm q} \\
0 & -s^{}_{\rm q} & c^{}_{\rm q} \\
\end{matrix}\right) \cdot
\left(
\begin{matrix}
c^{}_{\rm d} & -s^{}_{\rm d} & 0 \\
s^{}_{\rm d} & c^{}_{\rm d} & 0 \\
0 & 0 & 1 \\
\end{matrix}\right) \; ,
\label{eq:FX}
\end{eqnarray}
with $s^{}_{\rm u,q,d} \equiv \sin \theta^{}_{\rm u,q,d}$ and  $c^{}_{\rm u,q,d} \equiv \cos \theta^{}_{\rm u,q,d}$. Inspired by the FX parametrization, we assume $V^{\rm L}_{\rm u}$ and $V^{\rm L}_{\rm d}$ to be the products of $(2,3)$- and $(1,2)$-rotations, i.e.,
\begin{eqnarray}
V^{\rm L}_{\rm u} = \left(
\begin{matrix}
1 & 0 & 0 \\
0 & c^{}_{\rm t} & s^{}_{\rm t} \\
0 & -s^{}_{\rm t} & c^{}_{\rm t} \\
\end{matrix}\right) \cdot
\left(
\begin{matrix}
c^{}_{\rm u} & -s^{}_{\rm u} & 0 \\
s^{}_{\rm u} & c^{}_{\rm u} & 0 \\
0 & 0 & 1 \\
\end{matrix}\right) \; , \quad
V^{\rm L}_{\rm d} = \left(
\begin{matrix}
e^{\rm -i\phi} & 0 & 0 \\
0 & c^{}_{\rm b} & s^{}_{\rm b} \\
0 & -s^{}_{\rm b} & c^{}_{\rm b} \\
\end{matrix}\right) \cdot
\left(
\begin{matrix}
c^{}_{\rm d} & -s^{}_{\rm d} & 0 \\
s^{}_{\rm d} & c^{}_{\rm d} & 0 \\
0 & 0 & 1 \\
\end{matrix}\right) \; .
\label{eq:ULDL}
\end{eqnarray}
Then one can immediately find that the CKM matrix has exactly the same form as Eq.~(\ref{eq:FX}) by redefining $\theta^{}_{\rm q} \equiv \theta^{}_{\rm b}-\theta^{}_{\rm t}$. It is necessary to mention that there are totally five parameters in Eq.~(\ref{eq:ULDL}), which is one more than the number of parameters in Eq.~(\ref{eq:FX}).  As we will see later, the additional small rotation angle $\theta^{}_{\rm t}$ in Eq.~(\ref{eq:ULDL}) is requisite to give the correct contribution to $R^{}_{K^{(\ast)_{}}}$. For the right-handed quark fields, $V^{\rm R}_{\rm u}$ and $V^{\rm R}_{\rm d}$ are simply the identity matrices. As a result, there will be no flavor-changing neutral currents in the right-handed quark sector.

Given the unitary matrices $V^{\rm L,R}_{\rm u}$ and $V^{\rm L,R}_{\rm d}$, now we can reconstruct the mass matrices of up- and down-type quarks
\begin{eqnarray}
M^{}_{\rm u} \approx \left(
\begin{matrix}
c^{}_{\rm u} m^{}_{u} & -s^{}_{\rm u} m^{}_c & 0 \\
s^{}_{\rm u}c^{}_{\rm t} m^{}_{u} & c^{}_{\rm u}c^{}_{\rm t}m^{}_{c} & s^{}_{\rm t}m^{}_{t} \\
0 & -c^{}_{\rm u}s^{}_{\rm t}m^{}_{c} & c^{}_{\rm t} m^{}_{t} \\
\end{matrix}\right) \; , \quad
M^{}_{\rm d} \approx \left(
\begin{matrix}
e^{-{\rm i}\phi}_{}c^{}_{\rm d} m^{}_{d} & -e^{-{\rm i}\phi}_{}s^{}_{\rm d} m^{}_{s} & 0 \\
s^{}_{\rm d} c^{}_{\rm b}  m^{}_{d} & c^{}_{\rm d}c^{}_{\rm b}m^{}_{s} & s^{}_{\rm b}m^{}_{b} \\
0 & -c^{}_{\rm d}s^{}_{\rm b}m^{}_{s} & c^{}_{\rm b}m^{}_{b} \\
\end{matrix}\right) \; ,
\label{eq:Mqrec}
\end{eqnarray}
where the approximate relations $s^{}_{\rm u,d}s^{}_{\rm t,b}m^{}_{u,d} \approx 0$ have been adopted. Eq.~(\ref{eq:Mqrec}) implies all the free parameters except $y^{\rm d}_{11}$ and $y^{\rm d}_{12}$ in Eq.~(\ref{eq:mass_quark}) should be real. Meanwhile, from Eq.~(\ref{eq:Mqrec}) one can obtain that
\begin{eqnarray}
\left|\frac{y^{\rm u,d}_{21}}{y^{\rm u,d}_{11}}\right| =  \cos\theta^{}_{\rm t,b}\tan\theta^{}_{\rm u,d} \; , \quad \left|\frac{y^{\rm u,d}_{12}}{y^{\rm u,d}_{22}}\right| =  \frac{\tan\theta^{}_{\rm u,d}}{\cos\theta^{}_{\rm t,b}}\; , \quad \frac{\widetilde{y}^{\rm u,d}_{32}}{y^{\rm u,d}_{22}} = -\frac{\widetilde{y}^{\rm u,d}_{23}}{y^{\rm u,d}_{33}} = -\frac{\tan \theta^{}_{\rm t,b}}{\xi} \; . 
\label{eq:quark_rel}
\end{eqnarray}
The best-fit values of magnitudes of all nine CKM elements from Ref.~\cite{PDG2020} are
\begin{eqnarray}
V^{\rm bf}_{} = \left(
\begin{matrix}
0.97446 && 0.22452 && 0.00365 \\
0.22438 && 0.97359 && 0.04214 \\
0.00896 && 0.04133 && 0.999105 \\ 
\end{matrix}\right) \; ,
\label{eq:ckmbf}
\end{eqnarray}
which lead to the best-fit values of $\{\theta^{}_{\rm u},\theta^{}_{\rm d}, \theta^{}_{\rm q}, \phi\}$ as
\begin{eqnarray}
\theta^{}_{\rm u} =  0.0867\; , \quad \theta^{}_{\rm q} = 0.0423 \; , \quad \theta^{}_{\rm d} =  0.2148 \; , \quad \phi = 87.08^\circ_{} \; .
\label{eq:angle_bf}
\end{eqnarray}
Then Eq.~(\ref{eq:quark_rel}) indicates that the ratios $|y^{\rm u,d}_{21}/y^{\rm u,d}_{11}|$ and $|y^{\rm u,d}_{12}/y^{\rm u,d}_{22}|$ are suppressed respectively by the small values of $\theta^{}_{\rm u}$ and $\theta^{}_{\rm d}$, while $\widetilde{y}^{\rm u,d}_{32}$ ($\widetilde{y}^{\rm u,d}_{23}$) and $y^{\rm u,d}_{22}$ ($y^{\rm u,d}_{33}$) could be comparable if $\xi$ and $\theta^{}_{\rm t,b}$ are of the same order. We should mention that it is difficult to determine the general forms of $V^{\rm L,R}_{\rm u}$ and $V^{\rm L,R}_{\rm d}$ because of the redundant parameters in $M^{}_{\rm u,d}$. On this account, instead of implementing the standard diagonalization of $M^{}_{\rm u,d}$, we assume the specific forms of $V^{\rm L,R}_{\rm u}$ and $V^{\rm L,R}_{\rm d}$ {\it a priori} by observing the structures of $M^{}_{\rm u,d}$, and figure out the conditions under which our assumptions can be realistic. It is found that if the relations given in Eq.~(\ref{eq:quark_rel}) are satisfied, $V^{\rm L}_{\rm u,d}$ indeed take the forms shown in Eq.~(\ref{eq:ULDL}) while  $V^{\rm R}_{\rm u,d}$ are diagonal. 

Now we turn to the lepton sector. Since the charged-lepton mass matrix is diagonal, the lepton flavor mixing originates solely from the effective neutrino mass matrix, which takes the ${\bf A}^{}_2$ two-zero texture. Diagonalizing $M^{}_\nu$ via $U^\dag_{\nu} M^{}_{\nu} U^{\ast}_\nu = {\rm Diag}\{m^{}_1, m^{}_2, m^{}_3\}$ with $m^{}_i$ (for $i=1,2,3$) being three light neutrino masses, one can obtain the Pontecorvo-Maki-Nakagawa-Sakata (PMNS) matrix~\cite{Pontecorvo:1957cp,Maki:1962mu} $U = U^{}_\nu$. The phenomenological aspects of ${\bf A}^{}_2$ for neutrino masses and lepton flavor mixing have been investigated in detail in Refs.~\cite{Fritzsch:2011qv,Zhou:2015qua}. Up to the first order of $\sin\theta^{}_{13}$, we have~\cite{Fritzsch:2011qv,Zhou:2015qua}
\begin{eqnarray}
\dfrac{m^{}_{1}}{m^{}_{3}} \approx \tan \theta^{}_{12} \cot \theta^{}_{23} \sin \theta^{}_{13} \; , \quad
\dfrac{m^{}_{2}}{m^{}_{3}} \approx \cot \theta^{}_{12} \cot \theta^{}_{23} \sin \theta^{}_{13} \; ,
\label{eq:A2app1}
\end{eqnarray}
where $\theta^{}_{ij}$ (for $ij=12,13,23$) are three mixing angles of $U$ in the standard parametrization~~\cite{PDG2020}. The best-fit values of three mixing angles from the global-fit analysis by NuFIT~5.0~\cite{Esteban:2020cvm,nufit5.0} are $\theta^{}_{12} \approx 33^\circ_{}$, $\theta^{}_{13} \approx 8.6^\circ_{}$ and $\theta^{}_{23} \approx 49^\circ_{}$, hence it is easy to identify that Eq.~(\ref{eq:A2app1}) leads to $m^{}_1<m^{}_2<m^{}_3$, i.e., only the normal mass ordering (NO) is favored by ${\bf A}^{}_2$. The Dirac CP-violating phase $\delta$ and two Majorana CP-violating phases $\rho$ and $\sigma$ approximate to be\cite{Fritzsch:2011qv}\footnote{In this paper, we follow the convention that the Majorana phase matrix is denoted as $P={\rm Diag}\{e^{{\rm i}\rho}_{},e^{{\rm i}\sigma}_{},1\}$.}
\begin{eqnarray}
\cos\delta &\approx& \frac{\cot \theta^{}_{23}}{\tan 2 \theta^{}_{12} \sin \theta^{}_{13}}\left(1-\frac{\sin 2 \theta^{}_{12} \tan 2 \theta^{}_{12} R^{}_{\nu}}{4 \cot^{2}_{} \theta^{}_{23} \sin^{2}_{} \theta^{}_{13}}\right) \; , \nonumber \\
\rho & \approx & \frac{\pi}{2}-\frac{\delta}{2} \; , \nonumber \\
\sigma & \approx & \pi - \frac{\delta}{2} \; ,
\label{eq:A2app2}
\end{eqnarray}
where $R^{}_\nu \equiv \Delta m^2_{21}/\Delta m^2_{31}$ with $\Delta m^2_{21} \equiv m^2_2-m^2_1$ and $\Delta m^2_{31} \equiv m^2_3-m^2_1$ being two neutrino mass-squared differences has been defined. Substituting the best-fit values of three mixing angles $\theta^{}_{ij}$ and the ratio $R^{}_\nu \approx 0.0295$ into the first equation of Eq.~(\ref{eq:A2app2}), we arrive at $\cos\delta \approx 0.133$, i.e., $\delta \approx 278^\circ_{}$. Therefore, the ${\bf A}^{}_2$ two-zero texture tends to predict relatively large CP violation. The last two equations of Eq.~(\ref{eq:A2app2}) imply $\sigma - \rho \approx \pi/2$, which remains to be tested in the following numerical analysis.
\begin{figure}[t!]
	\centering		\includegraphics[width=1\textwidth]{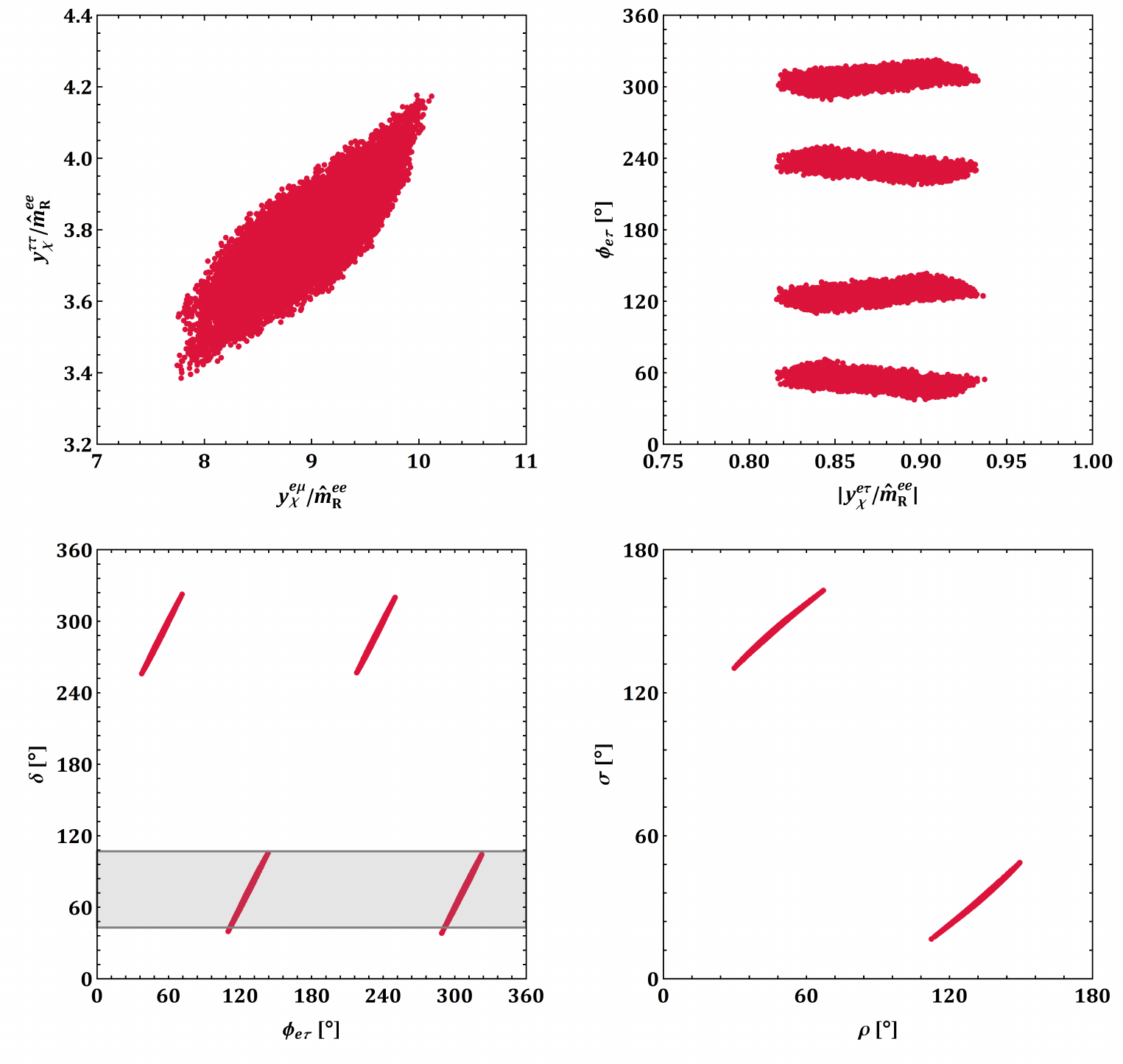} 	\vspace{0cm}
	\caption{The allowed parameter space of the model parameters $\{y^{e\mu}_\chi/\widehat{m}^{ee}_{\rm R},y^{\tau\tau}_\chi/\widehat{m}^{ee}_{\rm R},|y^{e\tau}_\chi/\widehat{m}^{ee}_{\rm R}|,\phi^{}_{e\tau}\}$ at the $1\sigma$ level and the predictions for three CP-violating phases in the NO case, where the gray shaded area in the bottom-left panel represents the excluded region inconsistent with the $3\sigma$ allowed range of $\delta$ from the global-fit results~\cite{Esteban:2020cvm,nufit5.0}.}
	\label{fig:figPara} 
	\vspace{0.6cm}
\end{figure}

As can be seen in Eq.~(\ref{eq:mass_lepton}), there are three real parameters $y^\nu_\alpha$ (for $\alpha = e, \mu, \tau$) in the Dirac neutrino mass matrix $M^{}_{\rm D}$, and three real parameters $\{\widehat{m}^{ee}_{\rm R}, y^{e\mu}_\chi, y^{\tau\tau}_\chi\}$ together with one complex parameter $y^{e\tau}_\chi$ in $M^{}_{\rm R}$. However, not all these parameters are independent. With out loss of generality, we assume three Dirac neutrino Yukawa coupling constants are equal to each other, i.e., $y^\nu_e = y^\nu_\mu = y^\nu_\tau$. Then there are only four real parameters apart from an overall factor in the effective neutrino mass matrix $M^{}_\nu$. In order to find out the allowed parameter space of these parameters, we take the $1\sigma$ ranges of two neutrino mass-squared differences and three mixing angles in the NO case from the NuFIT~5.0 results without including the atmospheric neutrino data from Super-Kamiokanda~\cite{Esteban:2020cvm,nufit5.0}, namely,
\begin{eqnarray}
\begin{array}{c}
\sin^2_{}\theta^{}_{12} = 0.304^{+0.013}_{-0.012} \; , \quad \sin^2_{}\theta^{}_{13} = 0.02221^{+0.00068}_{-0.00062} \; , \quad \sin^2_{}\theta^{}_{23} = 0.570^{+0.018}_{-0.024} \; , \\
\Delta m^{2}_{21} = \left(7.42^{+0.21}_{-0.20}\right) \times 10^{-5}_{}~{\rm eV} \; , \quad
\Delta m^{2}_{31} = \left(2.514^{+0.028}_{-0.027}\right) \times 10^{-3}_{}~{\rm eV} \; .
\end{array}
\label{eq:bfval}
\end{eqnarray}
We randomly generate the values of model parameters $\{y^{e\mu}_\chi/\widehat{m}^{ee}_{\rm R},y^{\tau\tau}_\chi/\widehat{m}^{ee}_{\rm R},|y^{e\tau}_\chi/\widehat{m}^{ee}_{\rm R}|\}$ and $\phi^{}_{e\tau}$ within the ranges $[10^{-4}_{},10^4_{}]$ and $[0,2\pi]$, respectively, and calculate the predictions for the above low-energy observables by numerically diagonalizing $M^{}_\nu$ in Eq.~(\ref{eq:mass_neutrino}). Then we compare these predictions with the ranges listed in Eq.~(\ref{eq:bfval}), and obtain the $1\sigma$ allowed parameter space of the model parameters. 

The numerical results are displayed in Fig.~\ref{fig:figPara}. From the top two panels of Fig.~\ref{fig:figPara} we can observe that $|y^{e\tau}_\chi|$, $y^{\tau\tau}_{\chi}$ and $\widehat{m}^{ee}_{\rm R}$ are essentially of the same order if $\xi = 0.1$ is assumed, whereas $y^{e\mu}_{\chi}$ is about ten times larger than $\widehat{m}^{ee}_{\rm R}$. There are separate ranges in the allowed parameter space of $\phi^{}_{e\tau}$. Nevertheless, as can be seen in the bottom-left panel of Fig.~\ref{fig:figPara}, only two of them with $36^\circ_{} < \phi^{}_{e\tau} < 72^\circ_{} $ and $216^\circ_{} < \phi^{}_{e\tau}< 252^\circ_{}$ will be retained if the $3\sigma$ allowed range of $\delta$ from the global-fit results~\cite{Esteban:2020cvm,nufit5.0} is taken into consideration. In the bottom-right panel, we also exhibit the correlation between two Majorana CP-violating phases $\rho$ and $\sigma$, where one can confirm that $\sigma - \rho \approx \pi/2$ is indeed a good approximation.

In short, by implementing an anomaly-free $U(1)^{}_{B^{}_3-2L^{}_\mu/3-L^{}_\tau/3}$ gauge symmetry on the canonical type-I seesaw mechanism, and allowing this symmetry to be spontaneously broken by an extra flavon field $\chi$, we obtain the mass matrices of quarks and leptons, which can successfully account for the observed fermion masses and flavor mixing. To be specific, in the quark sector both $V^{\rm L}_{\rm u}$ and $V^{\rm L}_{\rm d}$ can be ascribed to the combinations of $(2,3)$- and $(1,2)$-rotations, while $V^{\rm R}_{\rm u}$ and $V^{\rm R}_{\rm d}$ are simply the identity matrices. In the lepton sector, the effective neutrino mass matrix turns out to be the ${\bf A}^{}_2$ two-zero texture, which predicts the NO of three light neutrinos and relatively large CP violation.

\section{Solution to flavor anomalies}\label{sec:sol}
\subsection{The $R^{}_{K^{(\ast)_{}}}$ anomaly}
The flavor-changing process $b \rightarrow s\mu^+_{}\mu^-_{}$ can be described by the following effective Hamiltonian
\begin{eqnarray}
{\cal H}^{}_{\rm eff} = -\frac{4G^{}_{\rm F}}{\sqrt{2}}V^{\ast}_{tb}V^{}_{ts}(C^\mu_9{\cal O}^\mu_9+C^\mu_{10}{\cal O}^\mu_{10} ) \; ,
\label{eq:effHal}
\end{eqnarray}
where
\begin{eqnarray}
{\cal O}^\mu_9 = \frac{\alpha^{}_{\rm em}}{4\pi}(\overline{s}\gamma^\mu_{}b^{}_{\rm L})(\overline{\mu}\gamma_\mu^{}\mu) \; , \quad {\cal O}^\mu_{10} = \frac{\alpha^{}_{\rm em}}{4\pi}(\overline{s}\gamma^\mu_{} b^{}_{\rm L})(\overline{\mu}\gamma_\mu^{} \gamma^5_{} \mu) \; ,
\label{eq:Wilson_def}
\end{eqnarray}
with $G^{}_{\rm F}$ and $\alpha^{}_{\rm em} \equiv e^2_{}/(4\pi)$ being respectively the Fermi and fine-structure constants, and $C^\mu_{9}$ and $C^\mu_{10}$ are the corresponding Wilson coefficients. To see how the $Z^\prime_{}$ gauge boson in our model triggers the $b \rightarrow s$ transition process, let us rewrite Eq.~(\ref{eq:NC}) in the mass basis of quarks
\begin{eqnarray}
{\cal L}^{}_{\rm NC} &= & -\frac{g^{}_{Z^{\prime}_{}}}{3}\left(2\overline{\ell^{}_{\mu \mathrm{L}}} \gamma^{\mu}_{} \ell^{}_{\mu \mathrm{L}}+\overline{\ell^{}_{\tau \mathrm{L}}} \gamma^{\mu}_{} \ell^{}_{\tau \mathrm{L}}+2\overline{\mu^{}_{\mathrm{R}}} \gamma^{\mu}_{} \mu^{}_{\mathrm{R}}+\overline{\tau^{}_{\mathrm{R}}} \gamma^{\mu}_{} \tau^{}_{\mathrm{R}}+2\overline{\nu^{}_{\mu \mathrm{R}}} \gamma^{\mu}_{} \nu^{}_{\mu \mathrm{R}}+\overline{\nu^{}_{\tau \mathrm{R}}} \gamma^{\mu}_{} \nu^{}_{\tau \mathrm{R}}\right) Z_{\mu}^{\prime} \nonumber \\
&&+\frac{g^{}_{Z^{\prime}_{}}}{3}\left(\Gamma^{\rm u}_{ij}\overline{u^{\prime}_{i{\rm L}}}\gamma^\mu_{}u^{\prime}_{j{\rm L}} + \Gamma^{\rm d}_{ij}\overline{d^{\prime}_{i{\rm L}}}\gamma^\mu_{}d^{\prime}_{j{\rm L}}+\overline{t^{\prime}_{\rm R}}\gamma^\mu_{} t^{\prime}_{\rm R}  +\overline{b^{\prime}_{\rm R}}\gamma^\mu_{} b^{\prime}_{\rm R}\right) Z^\prime_\mu \; ,
\label{eq:NCmass}
\end{eqnarray}
where the superscript ``\,${}^\prime_{}$\,'' denotes the mass eigenstate and $\Gamma^{\rm u,d}_{}$ are defined as
\begin{eqnarray}
\Gamma^{\rm u}_{} \equiv \left(
\begin{matrix}
s^2_{\rm u}s^2_{\rm t} & c^{}_{\rm u}s^{}_{\rm u}s^2_{\rm t} & -s^{}_{\rm u} c^{}_{\rm t}s^{}_{\rm t}\\
c^{}_{\rm u}s^{}_{\rm u}s^2_{\rm t} & c^2_{\rm u}s^2_{\rm t} & -c^{}_{\rm u}c^{}_{\rm t}s^{}_{\rm t} \\
-s^{}_{\rm u}c^{}_{\rm t}s^{}_{\rm t} & -c^{}_{\rm u}c^{}_{\rm t}s^{}_{\rm t} & c^2_{\rm t}
\end{matrix}\right) \; , \quad 
\Gamma^{\rm d}_{} \equiv \left(
\begin{matrix}
s^2_{\rm d}s^2_{\rm b} & c^{}_{\rm d}s^{}_{\rm d}s^2_{\rm b} & -s^{}_{\rm d}c^{}_{\rm b}s^{}_{\rm b} \\
c^{}_{\rm d}s^{}_{\rm d}s^2_{\rm b} & c^2_{\rm d}s^2_{\rm b} & -c^{}_{\rm d}c^{}_{\rm b}s^{}_{\rm b} \\
-s^{}_{\rm d}c^{}_{\rm b}s^{}_{\rm b} & -c^{}_{\rm d}c^{}_{\rm b}s^{}_{\rm b} & c^2_{\rm b}
\end{matrix}\right) \; .
\label{eq:Gamma}
\end{eqnarray}

Now that the flavor-changing neutral-current Lagrangian is given, it is straightforward to integrate out the heavy $Z^\prime_{}$ gauge boson and derive the effective Lagrangian for the $b \rightarrow s\mu^+_{}\mu^-_{}$ process, namely,
\begin{eqnarray}
\Delta {\cal L}^{\rm NC}_{\rm eff} &=& -\frac{g^2_{Z^\prime_{}}\cos\theta^{}_{\rm d}\sin 2\theta^{}_{\rm b}}{9m^2_{Z^\prime_{}}} (\overline{s}^{}_{\rm L}\gamma^\mu_{}b^{}_{\rm L})(\overline{\mu}\gamma_\mu^{}\mu) + {\rm h.c.} \nonumber \\
&=& -\frac{g^2_{Z^\prime_{}}\cos\theta^{}_{\rm d}\sin 2\theta^{}_{\rm b}}{9m^2_{Z^\prime_{}}V^{}_{tb}V^{\ast}_{ts}} \left(\frac{\sqrt{2}\pi}{G^{}_{\rm F}\alpha^{}_{\rm em}}\right)\left(\frac{4G^{}_{\rm F}}{\sqrt{2}}\frac{\alpha^{}_{\rm em}}{4\pi}V^{}_{tb}V^{\ast}_{ts}\right) (\overline{s}^{}_{\rm L}\gamma^\mu_{}b^{}_{\rm L})(\overline{\mu}\gamma_\mu^{}\mu) + {\rm h.c.} \; ,
\label{eq:effL}
\end{eqnarray}
where one can extract contributions from the $Z^\prime_{}$ gauge boson to the Wilson coefficients $\Delta C^\mu_9$ and $\Delta C^\mu_{10}$ as
\begin{eqnarray}
\Delta C^\mu_9 = -\frac{\sqrt{2}\pi}{G^{}_{\rm F}\alpha^{}_{\rm em}} \frac{g^2_{Z^\prime_{}}\cos\theta^{}_{\rm d}\sin 2\theta^{}_{\rm b}}{9m^2_{Z^\prime_{}}V^{}_{tb}V^{\ast}_{ts}} \; , \quad \Delta C^\mu_{10} = 0 \; .
\label{eq:Wilson}
\end{eqnarray}
The $1\sigma$ allowed range of $\Delta C^\mu_9$ from the global-fit results under the condition that $\Delta C^\mu_{10} = 0$ in Ref.~\cite{Altmannshofer:2021qrr} yields $-0.94 < \Delta C^\mu_9 < -0.66$. Since $V^{}_{tb}V^\ast_{ts} = -\cos\theta^{}_{\rm d}\sin 2\theta^{}_{\rm q}/2$ in the FX parametrization, $\theta^{}_{\rm b}$ and $\theta^{}_{\rm q}$ should possess opposite signs so that we can arrive at $\Delta C^\mu_9 < 0$. This is the reason why we need an extra $(2,3)$-rotation in $V^{\rm L}_{\rm u}$. It is also worth noticing that in principle nonzero contributions to $\Delta C^\mu_{9,10}$ can be induced by the $S^{}_1$ leptoquark at the one-loop level. However, such contributions are negligibly small if we want to stay compatible with other experimental constraints, unless a relatively large leptoquark mass and specific coupling structures are assumed~\cite{Becirevic:2016oho,Cai:2017wry,Angelescu:2018tyl}. So we can safely neglect corrections from $S^{}_1$ to $R^{}_{K^{(\ast)}_{}}$, and consider $Z^\prime_{}$ as the only solution to the $R^{}_{K^{(\ast)}_{}}$ anomaly. 

\subsection{Muon $(g-2)$ and $R^{}_{D^{(\ast)_{}}}$ anomalies}
Although the $Z^\prime_{}$ model provides us with an excellent explanation of the $R^{}_{K^{(\ast)}_{}}$ anomaly, it can not induce the charged-current processes, and thus is unable to address the $R^{}_{D^{(\ast)_{}}}$ anomaly. On this account, in our work an $S^{}_1$ leptoquark plays the role of resolving the $R^{}_{D^{(\ast)_{}}}$ anomaly. Meanwhile, the $S^{}_1$ leptoquark also explains the observed muon $(g-2)$ result.

\begin{figure}[t!]
	\centering		\includegraphics[width=0.96\textwidth]{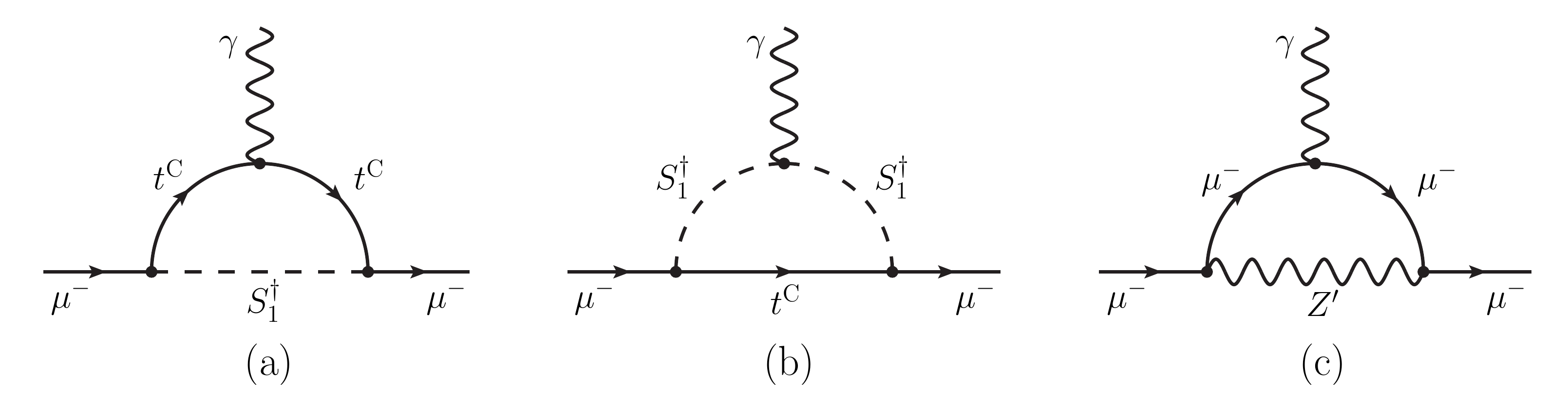} 	\vspace{-0.3cm}
	\caption{The Feynman diagrams for the dominant contributions from $S^{}_1$ and $Z^\prime_{}$ to the muon anomalous magnetic moment.}
	\label{fig:gmin2Feyn} 
	\vspace{0cm}
\end{figure}

Let us first consider the muon magnetic moment. The dominant contributions to the muon anomalous magnetic moment arising from $S^{}_1$ are illustrated in the first two Feynman diagrams in Fig.~\ref{fig:gmin2Feyn}. The corresponding correction to $\Delta a^{}_{\mu}$ is approximately given by~\cite{Bauer:2015knc,Chakraverty:2001yg,Cheung:2001ip}
\begin{eqnarray}
\Delta a^{S^{}_1}_\mu \approx \frac{m^{}_\mu m^{}_t {\rm Re}\,(\lambda^{\ell {\rm L}}_{3\mu} \lambda^{{\rm R}\ast}_{3\mu})}{4\pi^2_{}m^2_{S^{}_1}}\left(\log \frac{m^2_{S^{}_1}}{m^2_t}-\frac{7}{4}\right) \; ,
\label{eq:S1Damu}
\end{eqnarray}
where $\lambda^{\ell{\rm L}}_{} \equiv (V^{\rm L}_{\rm u})^{\rm T}_{}\lambda^{\rm L}_{}$ has been defined, and $m^{}_{S^{}_1}$ represents the mass of $S^{}_1$. Note that $m^{}_\mu \ll m^{}_t \ll m^{}_{S^{}_1}$ has been assumed in Eq.~(\ref{eq:S1Damu}). Apart from the $S^{}_1$ leptoquark, $Z^\prime_{}$ could also be the source of the muon anomalous magnetic moment [cf. diagram (c) in Fig.~\ref{fig:gmin2Feyn}]. Following the general formulas in Refs.~\cite{Leveille:1977rc,Moore:1984eg}, the modification to $\Delta a^{}_\mu$ from $Z^\prime_{}$ in our model can be expressed as
\begin{eqnarray}
\Delta a_{\mu}^{Z^{\prime}_{}}=\frac{g_{Z^{\prime}_{}}^{2} m_{\mu}^{2}}{9 \pi^{2}_{} m_{Z^{\prime}_{}}^{2}} \int_{0}^{1} \mathrm{d} x \frac{x^{2}_{}(1-x)}{(m_{\mu}^{2} / m_{Z^{\prime}_{}}^{2}) x^{2}_{}+(1-x)}  \; ,
\label{eq:ZDamu}
\end{eqnarray}
which approximates to be $\Delta a_{\mu}^{Z^{\prime}_{}} \approx (g_{Z^{\prime}_{}}^{2} m_{\mu}^{2})/(27 \pi^{2}_{} m_{Z^{\prime}_{}}^{2})$ in the region where $m^{}_\mu \ll m^{}_{Z^\prime_{}}$. However, the explanation of $R^{}_{K^{(\ast)}_{}}$ requires a heavy $Z^\prime_{}$ gauge boson with $m^{}_{Z^\prime_{}} \gtrsim 1~{\rm GeV}$, where the ratio $m^{}_{Z^\prime_{}}/g^{}_{Z^\prime_{}}$ is constrained to be larger than $550~{\rm GeV}$ by the neutrino trident process $\nu N \rightarrow \nu N \mu^+_{} \mu^-_{}$~\cite{Altmannshofer:2014pba}. Then it is not difficult to check that Eq.~(\ref{eq:ZDamu}) gives rise to $\Delta a_{\mu}^{Z^{\prime}_{}} \approx 15 \times 10^{-11}_{}$, which is much smaller than the observed value of $\Delta a^{}_\mu$. Hence it is reasonable to omit $\Delta a^{Z^\prime_{}}_\mu$ and regard $\Delta a^{S^{}_1}_\mu$ as the main contribution to $\Delta a^{}_\mu$.

The effective Lagrangian related to the charged-current process $b \rightarrow c\tau\nu$ can be written as
\begin{eqnarray}
{\cal L}^{\rm CC}_{\rm eff} &=& -\frac{4 G^{}_{\rm F}}{\sqrt{2}} V^{}_{c b}  \left[(\overline{c}^{}_{\rm L} \gamma^{}_{\mu} b^{}_{\rm L})(\overline{\tau}^{}_{\rm L} \gamma^{\mu}_{} \nu^{}_{\tau{\rm L}})+g_{\rm V}^{\ell}(\overline{c}^{}_{\rm L} \gamma^{}_{\mu} b^{}_{\rm L})(\overline{\tau}^{}_{\rm L} \gamma^{\mu}_{} \nu^{}_{\ell {\rm L}})+g_{\rm S}^{\ell}(\overline{c}^{}_{\rm R} b^{}_{\rm L}) (\overline{\tau}^{}_{\rm R} \nu^{}_{\ell {\rm L}})\right. \nonumber \\
&&\left.+g_{\rm T}^{\ell}(\overline{c}^{}_{\rm R} \sigma^{}_{\mu \nu} b^{}_{\rm L})(\overline{\tau}^{}_{\rm R} \sigma^{\mu \nu}_{} \nu_{\ell {\rm L}})\right]+ {\rm h.c. } \; ,
\label{eq:ccLag}
\end{eqnarray}
where the first term in the square brackets corresponds to the SM contribution, while the remaining terms come from the NP, with $g^{\ell}_{\rm V,S,T}$ (for $\ell = e, \mu, \tau$) being the Wilson coefficients. Notice that $g^\tau_{\rm V,S,T}$ can interfere with the SM coupling coefficient, whereas the other terms have no interference with the SM term. On this account, terms with $\ell = \tau$ in Eq.~(\ref{eq:ccLag}) dominate the NP contributions to the $b \rightarrow c\tau\nu$ process in most cases.

\begin{figure}[t!]
	\centering		\includegraphics[width=0.33\textwidth]{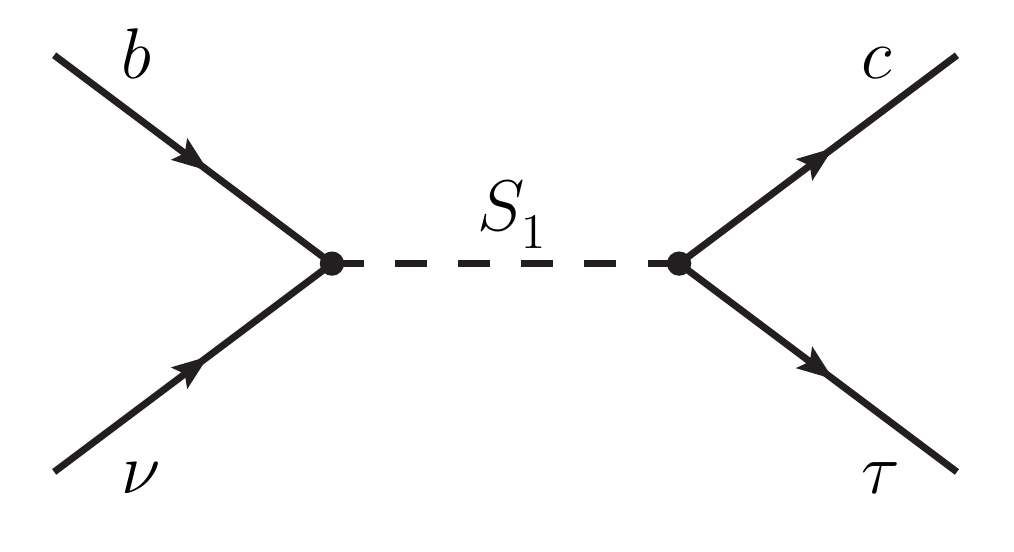} 	\vspace{-0.3cm}
	\caption{The Feynman diagram for the tree-level contribution from $S^{}_1$ to the $b \rightarrow c\tau\nu$.}
	\label{fig:cc} 
	\vspace{0cm}
\end{figure}

As for the $S^{}_1$ leptoquark, the tree-level contribution to the $b \rightarrow c\tau\nu$ process is demonstrated by the Feynman diagram exhibited in Fig.~\ref{fig:cc}. It is apparent that the non-zero Yukawa couplings among $S^{}_1$, $c^{}_{\rm L(R)}$ and $\tau^{}_{\rm L(R)}$ are requisite for the $b \rightarrow c\tau\nu$ process, leading to the equation $Q^{}_{\rm X} (S^{}_1) + x = 1$. This condition combined with the requirement that $S^{}_1$ should also couple to the $t$-quark as well as the muon gives the unique solution to $x$, namely, $x = 2/3$. This is the reason why we choose $X$ to be $B^{}_3 - 2L^{}_\mu/3 - L_\tau^{}/3$. Furthermore, such charge assignments will unavoidably forbid the coupling among $S^{}_1$, $b$ and $\nu^{}_\tau$ if we do not take the small CKM mixing into consideration. Instead, the term proportional to $\overline{b^{\rm C}_{\rm L}}\nu^{}_{\mu{\rm L}}S^{}_1$ will arise in the Lagrangian. Hence the dominant $b \rightarrow c\tau \nu$ decay mediated by the $S^{}_1$ leptoquark in our model is $b \rightarrow c\tau\nu^{}_\mu$, which does not interfere with the SM case. As a consequence, we should amplify the values of relevant leptoquark Yukawa couplings to achieve sizable contributions to $R^{}_{D^{(\ast)}_{}}$.

Another important point is that after integrating out the $S^{}_1$ leptoquark, the effective Wilson coefficients $g^{\ell}_{\rm V,S,T}$ in Eq.~(\ref{eq:ccLag}) are obtained at the energy scale $\mu^{}_{\rm R} = m^{}_{S^{}_1}$, which should then run down to the $b$-quark mass scale. The renormalization group (RG) running of the strong coupling constant $\alpha^{}_{\rm S}$ can provide considerable modifications to the Wilson coefficients $g^{\ell}_{\rm S,T}$, but $g^{\ell}_{\rm V}$ will not be affected by the RG running due to the Ward identity of QCD~\cite{Dorsner:2013tla,Gonzalez-Alonso:2017iyc}. In addition, finite corrections arising from the one-loop matching between the SM effective field theories (EFT) and the low-energy EFT~\cite{Dekens:2019ept} are also included. Assuming the mass of $S^{}_1$ to be $m^{}_{S^{}_1} \simeq 1~{\rm TeV}$, the Wilson coefficients at the scale $m^{}_b = 4.2~{\rm GeV}$ can be approximately expressed as~\cite{Gherardi:2020qhc}
\begin{eqnarray}
g^{\ell}_{\rm V} &\approx& + 1.09\frac{v^{2}_{H}}{2V^{}_{cb}}  \frac{\lambda_{3\ell}^{\nu {\rm L}} \lambda_{2 \tau}^{\ell {\rm L} \ast}}{2 m^2_{S^{}_1}} \; , \nonumber \\
g^{\ell}_{\rm S} &\approx& -1.63 \frac{v^{2}_H}{2 V^{}_{c b}} \frac{\lambda_{3 \ell}^{\nu {\rm L}} \lambda_{2 \tau}^{ {\rm R} \ast}}{2 m^2_{S^{}_1}}  \; ,\nonumber  \\
g^{\ell}_{\rm T} &\approx& +  0.88 \frac{v^{2}_H}{2 V^{}_{c b}} \frac{\lambda_{3 \ell}^{\nu {\rm L}} \lambda_{2 \tau}^{{\rm R} \ast}}{8 m^2_{S^{}_1}}  \; ,
\label{eq:WCRGE}
\end{eqnarray}
where we have defined $\lambda^{\nu{\rm L}}_{} \equiv (V^{\rm L}_{\rm d})^{\rm T}_{}\lambda^{\rm L}_{}$. It should be noticed that after transforming to the mass basis, the PMNS matrix may also appear in the definition of $\lambda^{\nu{\rm L}}_{}$. However, experimentally we are unable to distinguish different mass eigenstates of light neutrinos in the $b \rightarrow c\tau\nu$ process. In fact, in the basis where the charged-lepton mass matrix is diagonal, all the left-handed neutrino mass eigenstates $\nu^\prime_k$ are accompanied by the PMNS matrix elements $U^{}_{\ell k}$. When calculating the corresponding decay rates, we have already summed over the index $k$. As a consequence, we can equivalently regard neutrinos as massless particles and there is no PMNS mixing. 

With the help of Eq.~(\ref{eq:WCRGE}), now we can write down the approximate expressions of the observables mostly relevant to $b \rightarrow c\tau\nu$~\cite{Gherardi:2020qhc}
\begin{eqnarray}
\frac{R^{}_{D}}{R^{\rm SM}_{D}} &\approx& 1-0.79\frac{\lambda^{\nu{\rm L}\ast}_{3\tau}\lambda^{\rm R}_{2\tau}}{\widehat{m}^{2}_{S^{}_1}}+0.37\frac{(\lambda^{\nu{\rm L}\ast}_{3\mu}\lambda^{\rm R}_{2\tau})^2_{}}{\widehat{m}^{4}_{S^{}_1}} \; , \nonumber \\
\frac{R^{}_{D^\ast_{}}}{R^{\rm SM}_{D^\ast_{}}} &\approx& 1-0.34\frac{\lambda^{\nu{\rm L}\ast}_{3\tau}\lambda^{\rm R}_{2\tau}}{\widehat{m}^{2}_{S^{}_1}}+0.12\frac{(\lambda^{\nu{\rm L}\ast}_{3\mu}\lambda^{\rm R}_{2\tau})^2_{}}{\widehat{m}^{4}_{S^{}_1}} \; , \nonumber \\
\frac{{\cal B}(B^+_c \rightarrow \tau^+_{} \nu)}{{\cal B}(B^+_c \rightarrow \tau^+_{} \nu)^{\rm SM}_{}} &\approx& 1+5.1\frac{\lambda^{\nu{\rm L}\ast}_{3\tau}\lambda^{\rm R}_{2\tau}}{\widehat{m}^{2}_{S^{}_1}}+6.5\frac{(\lambda^{\nu{\rm L}\ast}_{3\mu}\lambda^{\rm R}_{2\tau})^2_{}}{\widehat{m}^{4}_{S^{}_1}} \; , 
\label{eq:obratio}
\end{eqnarray}
where $\widehat{m}^{}_{S^{}_1}$ is a dimensionless parameter defined as $\widehat{m}^{}_{S^{}_1} \equiv m^{}_{S^{}_1}/(1~{\rm TeV})$, and the SM predictions for $R^{}_{D}$, $R^{}_{D^\ast_{}}$ and ${\cal B}(B^+_c \rightarrow \tau^+_{} \nu)$ are~\cite{Amhis:2019ckw}
\begin{eqnarray}
R^{\rm SM}_{D} = 0.299 \; , \quad R^{\rm SM}_{D^{\ast}_{}} = 0.258 \; ,  \quad {\cal B}(B^+_c \rightarrow \tau^+_{} \nu)^{\rm SM}_{} = 0.023 \; .
\label{eq:RDSMbf}
\end{eqnarray}
The vector Wilson coefficients $g^{\ell}_{\rm V}$ have been omitted in Eq.~(\ref{eq:obratio}), due to the fact that  $\lambda_{2 \tau}^{\ell {\rm L}}$ should be small to escape the constraints from $B \rightarrow K^{(\ast)}_{}\nu\nu$ (cf. subsection~\ref{subsec:cons}). The terms proportional to $\lambda^{\nu{\rm L}\ast}_{3\tau}\lambda^{\rm R}_{2\tau}$ in Eq.~(\ref{eq:obratio}) are non-zero thanks to the non-trivial $V^{\rm L}_{\rm d}$, but their magnitudes are suppressed by the small value of $\theta^{}_{\rm b}$, and thus should be insignificant.

\subsection{Other constraints}\label{subsec:cons}
Before performing a concrete numerical analysis, we shall first collect other processes mediated by the $Z^\prime_{}$ gauge boson or the $S^{}_1$ leptoquark, e.g., the LFV decays, $B$- and $D$-meson decays, $B^{}_s-\overline{B}^{}_s$ mixing, $D^{0}_{}-\overline{D}{}^0_{}$ mixing, electroweak measurements, and the direct leptoquark searches at colliders. Experimental observations of these processes can impose important constraints on the allowed parameter space of our model. 

\subsubsection{LFV processes}
As the charged-lepton mass matrix is assumed to be diagonal, there will be no flavor-changing neutral-current interaction in the charged-lepton sector arising from the $Z^\prime_{}$ boson. Consequently, $Z^\prime_{}$ can not trigger the LFV decays in our model. However, the $S^{}_1$ leptoquark mediates the $\tau \rightarrow \mu\gamma$ process at the one-loop level, which receives the chirality enhancement and leads to the stringent constraint on our model. The branching ratio of $\tau \rightarrow \mu \gamma$ can be expressed as
\begin{eqnarray}
{\cal B}(\tau \rightarrow \mu\gamma) = \frac{m^3_\tau}{16\pi \Gamma^{}_\tau}\left(|T^{\rm L}_{\mu\tau}|^2_{}+|T^{\rm R}_{\mu\tau}|^2_{}\right) \; ,
\label{eq:brtmg}
\end{eqnarray}
where $\Gamma^{}_\tau$ is the decay width of $\tau$, and $T^{\rm L}_{\mu\tau}$ and $T^{\rm R}_{\mu\tau}$ are found to be~\cite{Dorsner:2016wpm}
\begin{eqnarray}
T^{\rm L}_{\mu\tau} &\approx& -\frac{e m^{}_t}{8\pi^2_{}}\frac{\lambda^{\ell {\rm L}}_{3\tau}\lambda^{\rm R \ast}_{3\mu}}{m^2_{S^{}_1}}\left(\log\frac{m^2_t}{m^2_{S^{}_1}} + \frac{7}{4}\right) \; , \nonumber \\
T^{\rm R}_{\mu\tau} &\approx& -\frac{e m^{}_c}{8\pi^2_{}}\frac{\lambda^{\ell {\rm L}\ast}_{2\mu}\lambda^{\rm R}_{2\tau}}{m^2_{S^{}_1}}\left(\log\frac{m^2_c}{m^2_{S^{}_1}} + \frac{7}{4}\right) \; .
\label{eq:Ttmg}
\end{eqnarray}
At first glance it seems that $T^{\rm L}_{\mu\tau}$ dominates the branching ratio of $\tau \rightarrow \mu\gamma$ given that $m^{}_t \gg m^{}_c$, which, however, is not the case since the combined explanations of $(g-2)$ and $R^{}_{D^{(\ast)}_{}}$ anomalies also require $\lambda^{\rm R}_{3\mu}$ to be much smaller than  $\lambda^{\rm R}_{2\tau}$, as will be seen in the numerical analysis later.

Apart from the $\tau \rightarrow \mu \gamma $ radiative decay, the four-lepton LFV process $\tau \rightarrow 3\mu$ can also be induced by $S^{}_1$ via the box diagram at the one-loop level. The leading term of ${\cal B}(\tau \rightarrow 3\mu)$ is proportional to the product of $|\lambda^{\ell{\rm L}}_{3\tau}|^2_{}$ and $|\lambda^{\ell{\rm R}}_{3\mu}|^{2}_{}$, both of which are restricted to be sufficiently small so that we will not consider the constraint from $\tau \rightarrow 3\mu$ on our model.

\subsubsection{$b \rightarrow c\mu(e)\nu$}
The $S^{}_1$ leptoquark coupling to the muon could modify the theoretical prediction for ${\cal B}(B \rightarrow D\mu \nu)$. Such a modification should be small to preserve the LFU between electrons and muons in the charge-current processes, which has been precisely examined in the experiments~\cite{BaBar:2008zui,Belle:2015pkj}. The LFU in $b \rightarrow c\mu(e)\nu$ can be embodied in the ratio
\begin{eqnarray}
R^{\mu/e}_{ D} \equiv \frac{{\cal B}(B \rightarrow D\mu\nu)}{{\cal B}(B \rightarrow De\nu)} \; ,
\label{eq:Rmue_def}
\end{eqnarray}
the dominant contribution of which from $S^{}_1$ reads~\cite{Gherardi:2020qhc}
\begin{eqnarray}
R^{\mu/e}_{D} \approx 1 +{\rm Re}\,\left(0.77 \frac{\lambda^{\nu {\rm L}}_{3\mu} \lambda^{\ell {\rm L}\ast}_{2\mu}}{V^{}_{cs}\widehat{m}^2_{S^{}_1}}\right) \; .
\label{eq:Rmue_app}
\end{eqnarray}
The combined results from BaBar~\cite{BaBar:2008zui} and Belle~\cite{Belle:2015pkj} experiments indicate that $R^{\mu/e}_{ D}|^{}_{\rm comb} = 0.978 \pm 0.035$. Since $\lambda^{\nu {\rm L}}_{3\mu} \approx \lambda^{\rm L}_{3\mu}$ and $\lambda^{\ell {\rm L}}_{2\mu} \approx \lambda^{\rm L}_{3\mu} \sin \theta^{}_{\rm t}$, the observed value of $R^{\mu/e}_{ D}$ can set tight upper bounds on $\lambda^{\rm L}_{3\mu}$ and $\theta^{}_{\rm t}$.

\subsubsection{$D^{}_s \rightarrow \tau\nu$}
Replacing the external line of $b$-quark in Fig.~\ref{fig:cc} by that of $s$-quark, we find that the leptonic decay $D^{}_s \rightarrow \tau\nu$ can take place at the tree level with $S^{}_1$ being the propagator. Taking the RG running effects into account, one arrives at~\cite{Gherardi:2020qhc}
\begin{eqnarray}
\frac{{\cal B}(D^{+}_s \rightarrow \tau^+_{}\nu)}{{\cal B}(D^{+}_s \rightarrow \tau^+_{}\nu)^{}_{\rm SM}} \approx 1 + 2\times 10^{-2}_{} {\rm Re}\,\left(1.5\frac{\lambda^{\nu {\rm L}\ast}_{2\tau}\lambda^{\ell {\rm L}}_{2\tau}}{V^{\ast}_{cs}\widehat{m}^2_{S^{}_1}} - 4.6 \frac{\lambda^{\nu {\rm L}\ast}_{2\tau}\lambda^{\rm R}_{2\tau}}{\widehat{m}^2_{S^{}_1}}\right) \; .
\label{eq:cstn}
\end{eqnarray}
The SM prediction for ${\cal B}(D^{+}_s \rightarrow \tau^+_{}\nu)$  is $(5.169 \pm 0.004) \times 10^{-2}_{}$~\cite{Aoki:2016frl}, and the experimental measurement on the $D^{+}_s \rightarrow \tau^+_{}\nu$ process gives ${\cal B}(D^{+}_s \rightarrow \tau^+_{}\nu) = (5.48 \pm 0.23) \times 10^{-2}_{}$~\cite{PDG2020}, yielding
\begin{eqnarray}
\frac{{\cal B}(D^{+}_s \rightarrow \tau^+_{}\nu)}{{\cal B}(D^{+}_s \rightarrow \tau^+_{}\nu)^{}_{\rm SM}} = 1.06 \pm 0.044 \; .
\label{eq:cstn_exp}
\end{eqnarray}

\subsubsection{$B \rightarrow K^{(\ast)}_{}\nu\nu$}
The rare meson decay $B \rightarrow K^{(\ast)}_{} \nu \nu$ can be mediated by both $Z^\prime_{}$ and $S^{}_1$, which is described by the following four-fermion interactions
\begin{eqnarray}
{\cal L}^{bs\nu\nu}_{\rm eff} = \frac{4G^{}_{\rm F}}{\sqrt{2}}V^\ast_{tb}V^{}_{ts}(C^{bs\alpha\beta}_{\nu,{\rm L}} {\cal O}^{bs\alpha\beta}_{\rm L}+C^{bs\alpha\beta}_{\nu,{\rm R}} {\cal O}^{bs\alpha\beta}_{\rm R})\; ,
\label{eq:bsnunuL}
\end{eqnarray}
with
\begin{eqnarray}
{\cal O}^{bs\alpha\beta}_{\rm L} &  = & \frac{\alpha^{}_{\rm em}}{4\pi}(\overline{s}^{}_{\rm L}\gamma^\mu_{}b^{}_{\rm L})[\overline{\nu}^{}_\alpha \gamma_\mu^{}(1-\gamma^{}_5)\nu^{}_\beta] \; , \nonumber \\
{\cal O}^{bs\alpha\beta}_{\rm R} &  = & \frac{\alpha^{}_{\rm em}}{4\pi}(\overline{s}^{}_{\rm R}\gamma^\mu_{}b^{}_{\rm R})[\overline{\nu}^{}_\alpha \gamma_\mu^{}(1-\gamma^{}_5)\nu^{}_\beta] \; .
\label{eq:bsnunuO}
\end{eqnarray}
At the tree level, the non-vanishing contributions come only from ${\cal O}^{bs\alpha\beta}_{\rm L}$. Therefore we can safely regard $C^{bs\alpha\beta}_{\nu,{\rm R}}$ as zero. The ratio $R^\nu_{K^{(\ast)}_{}} \equiv {\cal B}(B \rightarrow K^{(\ast)}_{}\nu\nu)/{\cal B}(B \rightarrow K^{(\ast)}_{}\nu\nu)^{}_{\rm SM}$ then is the function of Wilson coefficients  $C^{bs\alpha\beta}_{\nu,{\rm L}}$ and $C_{\nu}^{\mathrm{SM}}$, i.e.,
\begin{eqnarray}
R^{\nu}_{K^{(*)}}=\frac{1}{3}\sum^{}_{\alpha\beta}\frac{|C^{bs\alpha\beta}_{\nu,{\rm L}}|^{2}_{}}{|C_{\nu}^{\mathrm{SM}}|^{2}_{}} \; ,
\label{eq:RnuK_def}
\end{eqnarray}
where $C^{\rm SM}_\nu \approx -6.35$~\cite{Buras:2014fpa}. As for the $Z^\prime_{}$ boson, it is easy to derive that its correction to $R^{\nu}_{K^{(*)}}$ is $(5\Delta C^\mu_9)/(6 C^{\rm SM}_\nu)$ at the leading order, which is about 0.1 given that $\Delta C^\mu_9 \simeq -0.8$. The allowed ranges of $R^{\nu}_{K^{(*)}}$ at $95\%$ C.L. are~\cite{Belle:2017oht}
\begin{eqnarray}
R^\nu_K < 4.65 \; , \quad R^\nu_{K^{\ast}_{}} < 3.22 \; .
\label{RnuK_exp}
\end{eqnarray}
So we conclude that the modification from $Z^\prime_{}$ to $R^{\nu}_{K^{(*)}}$ is inappreciable. For the $S^{}_1$ leptoquark, the corresponding contribution to $R^{\nu}_{K^{(*)}}$ reads~\cite{Gherardi:2020qhc}
\begin{eqnarray}
R^{\nu}_{K^{(*)}} \approx 1 + {\rm Re}\, \left(1.37 \frac{\sum^{}_\alpha \lambda^{\nu{\rm L}\ast}_{2\alpha} \lambda^{\nu{\rm L}}_{3\alpha} }{|V^{}_{ts}|\widehat{m}^2_{S^{}_1}} + 1.42  \frac{\sum^{}_{\alpha\beta} |\lambda^{\nu{\rm L}\ast}_{2\alpha}|^2_{} |\lambda^{\nu{\rm L}}_{3\beta}|^2_{} }{|V^{}_{ts}|^2_{}\widehat{m}^4_{S^{}_1}} \right) \; ,
\label{eq:RnuK_app}
\end{eqnarray}
which leads to strong constraints on the rotation angle $\theta^{}_{\rm b}$ and the Yukawa couplings $\lambda^{\rm L}_{2\tau}$ and $\lambda^{\rm L}_{3\mu}$. 

\subsubsection{Meson mixing}
The $B^{}_{s}-\overline{B}^{}_s$ and $D^0_{}-\overline{D}{}^0_{}$ mixing processes receive NP corrections from the $Z^\prime_{}$ boson at the tree level. The Wilson coefficients relevant for these two processes are written as
\begin{eqnarray}
C^{Z^\prime_{}}_{B^{}_s} = \frac{g^2_{Z^\prime_{}}\cos^2_{}\theta^{}_{\rm d}\sin^2_{}2\theta^{}_{\rm b}}{72 m^2_{Z^\prime_{}}} \; , \quad C^{Z^\prime_{}}_{D} = \frac{g^2_{Z^\prime_{}}\sin^4_{}\theta^{}_{\rm t}\sin^2_{}2\theta^{}_{\rm u}}{72 m^2_{Z^\prime_{}}} \; ,
\label{eq:WilmixingZ}
\end{eqnarray}
respectively. $C^{Z^\prime_{}}_{B^{}_s}$ can be significant for large values of $g^{}_{Z^\prime_{}}/m^{}_{Z^\prime_{}}$ and $\theta^{}_{\rm b}$, thus provides stringent upper bounds on these two parameters. On the contrary, $C^{Z^\prime_{}}_{D}$  is suppressed by both $\sin^4_{}\theta^{}_{\rm t}$ and $\sin^2_{}2\theta^{}_{\rm u}$, so its constraint on our model is relatively weak.

Meanwhile, the $S^{}_1$ leptoquark induces the meson mixing at the one-loop level, leading to the following Wilson coefficients
\begin{eqnarray}
C^{S^{}_1}_{B^{}_s} = \frac{\sum^{}_\alpha(\lambda^{\nu{\rm L}\ast}_{3\alpha}\lambda^{\nu{\rm L}}_{2\alpha})^2_{}}{128\pi^2_{} m^2_{S^{}_1}} \; , \quad C^{S^{}_1}_{D} = \frac{\sum^{}_\alpha(\lambda^{\ell{\rm L}\ast}_{2\alpha}\lambda^{\ell{\rm L}}_{1\alpha })^2_{}}{128\pi^2_{} m^2_{S^{}_1}} \; ,
\label{eq:WilmixingS1}
\end{eqnarray}
both of which are suppressed by the small CKM mixing.~\footnote{It should be mentioned that $\lambda^{\ell {\rm L}}_{1\tau}$ in Eq.~(\ref{eq:WilmixingS1}) comes solely from the CKM mixing under the premise that we have omitted $\lambda^{\rm L}_{1\tau}$ in Eq.~(\ref{eq:LQ}). If $\lambda^{\rm L}_{1\tau}$ is nonzero, the contribution from $S^{}_1$ to $D^0_{}-\overline{D}{}^0_{}$ mixing could be considerable. In other words, the measurement of $D^0_{}-\overline{D}{}^0_{}$ mixing can restrict the magnitude of $\lambda^{\rm L}_{1\tau}$. For instance, if $\lambda^{\rm L}_{2\tau} \sim 0.1$, one will have $\lambda^{\rm L}_{1\tau} \lesssim 0.2$. } Therefore $S^{}_1$ can not make significant modifications to the $B^{}_{s}-\overline{B}^{}_s$ and $D^0_{}-\overline{D}{}^0_{}$ mixing, and will be neglected in the following numerical analysis.

\subsubsection{Precision electroweak measurements of $Z\ell\overline{\ell}$ couplings}
At the one-loop level, triangle diagrams with $S^{}_1$ inside the loop can modify the couplings $g^Z_{\alpha{\rm L}}$ and $g^Z_{\alpha{\rm R}}$ between the $Z$ gauge boson and leptons, which have been measured precisely in experiments. Corrections from $S^{}_1$ to the  $Z\ell\overline{\ell}$ couplings approximate to be~\cite{Marzocca:2021azj} 
\begin{eqnarray}
10^3_{}\delta g^Z_{\alpha {\rm L}} &\approx& 0.59\frac{|\lambda^{\ell{\rm L}}_{3\alpha}|^2_{}}{\widehat{m}^2_{S^{}_1}} \; , \nonumber \\
10^3_{}\delta g^Z_{\alpha\rm R} &\approx& -0.67 \frac{|\lambda^{\rm R}_{3\alpha}|^2_{}}{\widehat{m}^2_{S^{}_1}} + 0.06 \frac{|\lambda^{\rm R}_{2\alpha}|^2_{}}{\widehat{m}^2_{S^{}_1}} \; .
\label{eq:Zll}
\end{eqnarray}
All the elements of $\lambda^{\rm L}_{}$ and $\lambda^{\rm R}_{}$ except $\lambda^{\rm L}_{3\mu}$ and $\lambda^{\rm R}_{2\tau}$ are not sufficiently large to bring significant corrections to $Z\ell\overline{\ell}$ couplings. As a consequence, we will mainly focus on $\delta g^Z_{\mu {\rm L}}$ and $\delta g^Z_{\tau {\rm R}}$, the experimental constraints on which are~\cite{ALEPH:2005ab}
\begin{eqnarray}
\delta g^Z_{\mu {\rm L}} = (0.3 \pm 1.1) \times 10^{-3}_{} \; , \quad \delta g^Z_{\tau {\rm R}} = (0.66 \pm 0.65) \times 10^{-3}_{} \; .
\end{eqnarray}

\subsubsection{Collider constraints on the $S^{}_1$ leptoquark}
Leptoquarks can be produced at the LHC via pair production, resonant single production or the off-shell $t$-channel exchange in Drell-Yan processes~\cite{Dorsner:2018ynv,Schmaltz:2018nls, Buonocore:2020erb}. Leptoquarks from the pair production will decay into quarks and leptons, the branching ratios of which can set limits on the leptoquark masses~\cite{Angelescu:2018tyl,Schmaltz:2018nls}. For the $S^{}_1$ leptoquark, a lower bound on $m^{}_{S^{}_1}$ using the recent ATLAS and CMS data is found to be around $1~{\rm TeV}$ or less~\cite{Gherardi:2020qhc}. Different from the pair production, the leptoquark exchange in Drell-Yan processes relies also on the leptoquark Yukawa couplings to fermions~\cite{Angelescu:2018tyl,Schmaltz:2018nls,Faroughy:2016osc}, thus can be used to restrict the leptoquark Yukawa couplings. The latest $95\%$ C.L. upper limits on the Yukawa couplings $\{\lambda^{\rm L}_{2\tau}, \lambda^{\rm L} _{2\mu}, \lambda^{\rm R}_{2\tau}, \lambda^{\rm R} _{2\mu}\}$ using the most recent LHC searches in the high-$p^{}_{\rm T}$ bins of $pp \rightarrow \ell\ell$ at $13~{\rm TeV}$ with $140~{\rm fb}^{-1}_{}$ can be found in Ref.~\cite{Angelescu:2021lln}, while the limits on $\lambda^{\rm L}_{3\tau}$ and $\lambda^{\rm L}_{3\mu}$ obtained from $c\overline{c} \rightarrow \tau\overline{\tau},\mu\overline{\mu}$ are in the  non-perturbative region. In our model, $\lambda^{\rm L} _{2\mu}$ and $\lambda^{\rm R} _{2\mu}$ are exactly zero and $\lambda^{\rm L}_{2\tau}$ is also small, so we will only take the constraint on $\lambda^{\rm R}_{2\tau}$ into consideration. According to Ref.~\cite{Angelescu:2021lln}, we have $\lambda^{\rm R}_{2\tau} \lesssim 1.2$ for $m^{}_{S^{}_1} = 1~{\rm TeV}$.

\section{Numerical analysis}\label{sec:glo}
It is now the time to numerically test the feasibility of our model in explaining the muon anomalous magnetic moment and $B$-anomalies. We fix the values of $\{\theta^{}_{\rm u},\theta^{}_{\rm d}, \theta^{}_{\rm q}, \phi\}$ to be their individual best-fit values given in Eq.~(\ref{eq:angle_bf}), and the best-fit values of $\{m^{}_{c},m^{}_{t},m^{}_{\mu},m^{}_{\tau}\}$ are evaluated at the electroweak scale characterized by the mass of the $Z$ gauge boson $m^{}_{Z} \approx 91.2~{\rm GeV}$~\cite{Huang:2020hdv}, namely,
\begin{eqnarray}
m^{}_{c}=0.620 ~\mathrm{GeV}\; , \quad m^{}_{t}=168.26 ~\mathrm{GeV} \; , \quad
m^{}_{\mu}=0.101766 ~\mathrm{MeV} \; , \quad m^{}_{\tau}=1.72856 ~\mathrm{GeV} \; .
\label{eq:massbf}
\end{eqnarray}
Then the free parameters in our model can be summarized as: $\{m^{}_{Z^\prime_{}},g^{}_{Z^\prime_{}}\}$ associated with the $Z^\prime_{}$ boson, $\{m^{}_{S^{}_1},\lambda^{\rm L}_{2\tau},\lambda^{\rm L}_{3\mu},\lambda^{\rm R}_{2\tau},\lambda^{\rm R}_{3\mu}\}$ related to the $S^{}_1$ leptoquark, and an extra rotation angle $\theta^{}_{\rm b}$ (or equivalently, $\theta^{}_{\rm t} = \theta^{}_{\rm b}-\theta^{}_{\rm q}$). These parameters need to satisfy the experimental bounds of the observables listed in Table~\ref{table:expbound}. As we have mentioned before, the heavy $Z^\prime_{}$ gauge boson is sufficient to resolve the $R^{}_{K^{(\ast)_{}}}$ anomaly, while the $S^{}_1$  leptoquark can independently address the muon $(g-2)$ and  $R^{}_{D^{(\ast)_{}}}$ anomalies in our model. Furthermore, NP contributions to all the other observables discussed in subsection \ref{subsec:cons} rely dominantly on one of $Z^\prime_{}$ and $S^{}_1$. That is to say there is little correlation between the allowed parameter space of $\{m^{}_{Z^\prime_{}},g^{}_{Z^\prime_{}}\}$ and that of $\{m^{}_{S^{}_1},\lambda^{\rm L}_{2\tau},\lambda^{\rm L}_{3\mu},\lambda^{\rm R}_{2\tau},\lambda^{\rm R}_{3\mu}\}$, allowing us to investigate them separately.

\begin{table}[t!]
	\centering
	\caption{The best-fit values together with the $1\sigma$ intervals, or the upper bounds corresponding to $95\%$ C.L. of the relevant observables.}
	\vspace{0.5cm}
	\begin{tabular}{cc | c c }
		\hline\hline
		Observable & Experimental values&  Observable & Experimental values \\
		\hline
		$\Delta C^\mu_9$ &  $-0.8 \pm 0.14$~\cite{Altmannshofer:2021qrr} &  ${\rm Re}\,C^{}_{D}$ & $<3.57 \times 10^{-7}_{}~{\rm TeV}^{-2}_{}$~\cite{UTfit:2007eik}\\
		$\Delta a^{}_\mu$ & $(251 \pm 59) \times 10^{-11}_{}$~\cite{Abi:2021gix} & ${\rm Im}\,C^{}_{D}$ & $<2.23 \times 10^{-7}_{}~{\rm TeV}^{-2}_{}$~\cite{UTfit:2007eik}  \\
		$R^{}_D$ &  $ 0.34 \pm 0.029$~\cite{Amhis:2019ckw} & ${\cal B}(D^{+}_s \rightarrow \tau^+_{} \nu)$ & $(5.48 \pm 0.23) \times 10^{-2}_{}$~\cite{PDG2020} \\
		$R^{}_{D^{(\ast)}_{}}$ &  $0.295 \pm 0.013$~\cite{Amhis:2019ckw} & ${\cal B}(\tau \rightarrow \mu \gamma)$ & $<4.4 \times 10^{-8}_{}$~\cite{PDG2020} \\
		${\cal B}(B^+_c \rightarrow \tau^+_{}\nu)$  & $ < 0.1$~\cite{Akeroyd:2017mhr} & $ R^\nu_{K^{(\ast)}_{}}$ & $<3.22$~\cite{Belle:2017oht} \\
		$|C^{}_{B^{}_s}|$ & $<2.01 \times 10^{-5}_{}~{\rm TeV}^{-2}_{}$~~\cite{UTfit:2007eik} & $\delta g^Z_{\mu {\rm L}}$ & $(0.3 \pm 1.1) \times 10^{-3}_{}$~\cite{ALEPH:2005ab}  \\
		$R^{\mu/e}_{D}$ & $0.978 \pm 0.035$~\cite{BaBar:2008zui,Belle:2015pkj} & $\delta g^Z_{\tau {\rm R}}$ & $(0.66 \pm 0.65) \times 10^{-3}_{}$~\cite{ALEPH:2005ab}  \\
		\hline\hline
	\end{tabular}
	\label{table:expbound}
\end{table}

\begin{figure}[t!]
	\centering		\includegraphics[width=1\textwidth]{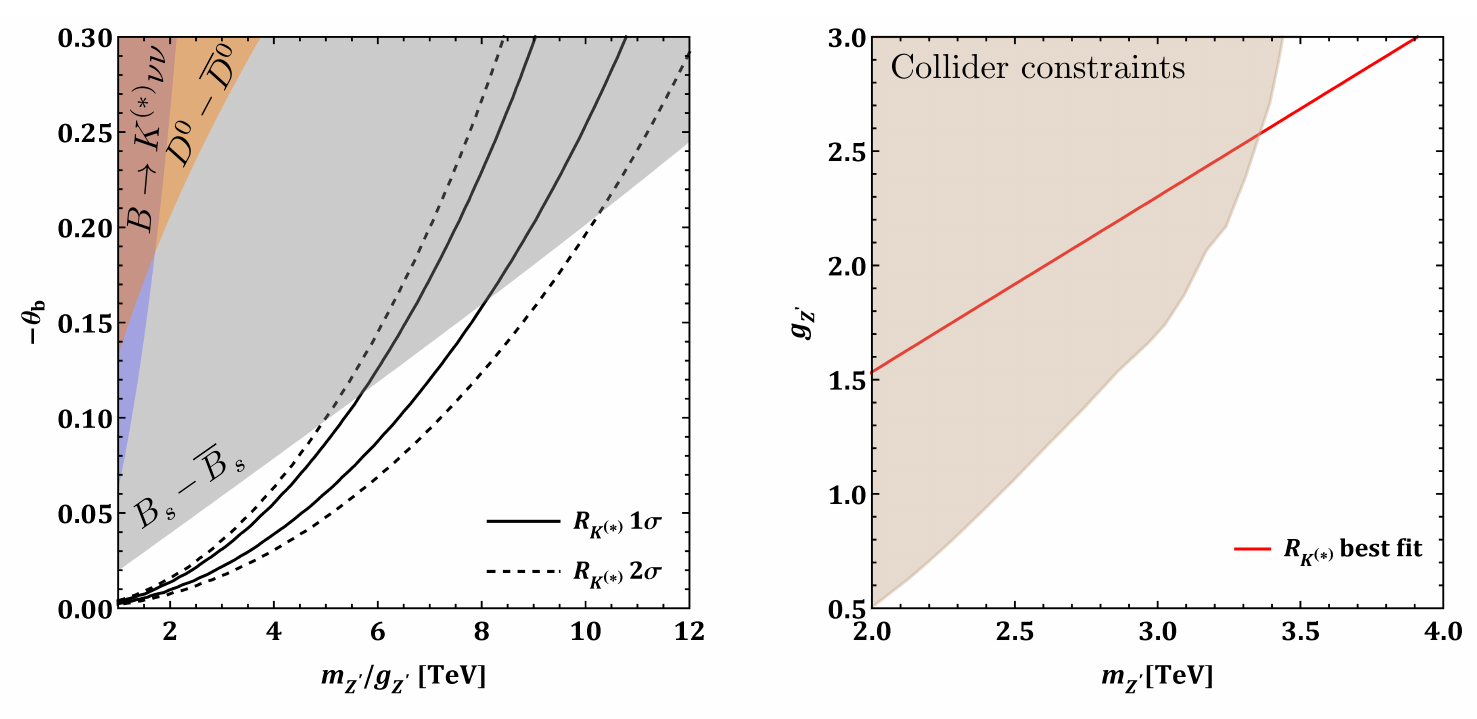} 	\vspace{-0.3cm}
	\caption{{\it Left panel}: The allowed parameter space of $\{m^{}_{Z^\prime_{}}/g^{}_{Z^\prime_{}},-\theta^{}_{\rm b}\}$ which is compatible with the observed $R^{}_{K^{(\ast)}_{}}$ anomaly at the $1\sigma$ (solid curves) and $2\sigma$ (dashed curves) levels. The gray-, orange- and blue-shaded regions are excluded by the $B^{}_s-\overline{B}^{}_s$ mixing, $D^{0}_{}-\overline{D}{}^0_{}$ mixing and $B \rightarrow K^{(*)}_{}\nu\nu$ at the $95\%$ C.L., respectively. {\it Right panel}: LHC constraints on the allowed parameters space of $m^{}_{Z^\prime_{}}$ and $g^{}_{Z^\prime_{}}$ by measuring the Drell-Yan process $q\overline{q} \to Z^\prime_{} \to \mu^+_{}\mu^{-}_{}$, where the value of $\theta^{}_{\rm b}$ is set to be $-0.005$. The brown region is excluded at the $95\%$ C.L. and the red line denotes the relation between $m^{}_{Z^\prime_{}}$ and $g^{}_{Z^\prime_{}}$ when $\Delta C^\mu_9$ takes its best-fit value.}
	\label{fig:C9} 
	\vspace{0cm}
\end{figure}

\begin{figure}[t!]
	\centering		\includegraphics[width=1\textwidth]{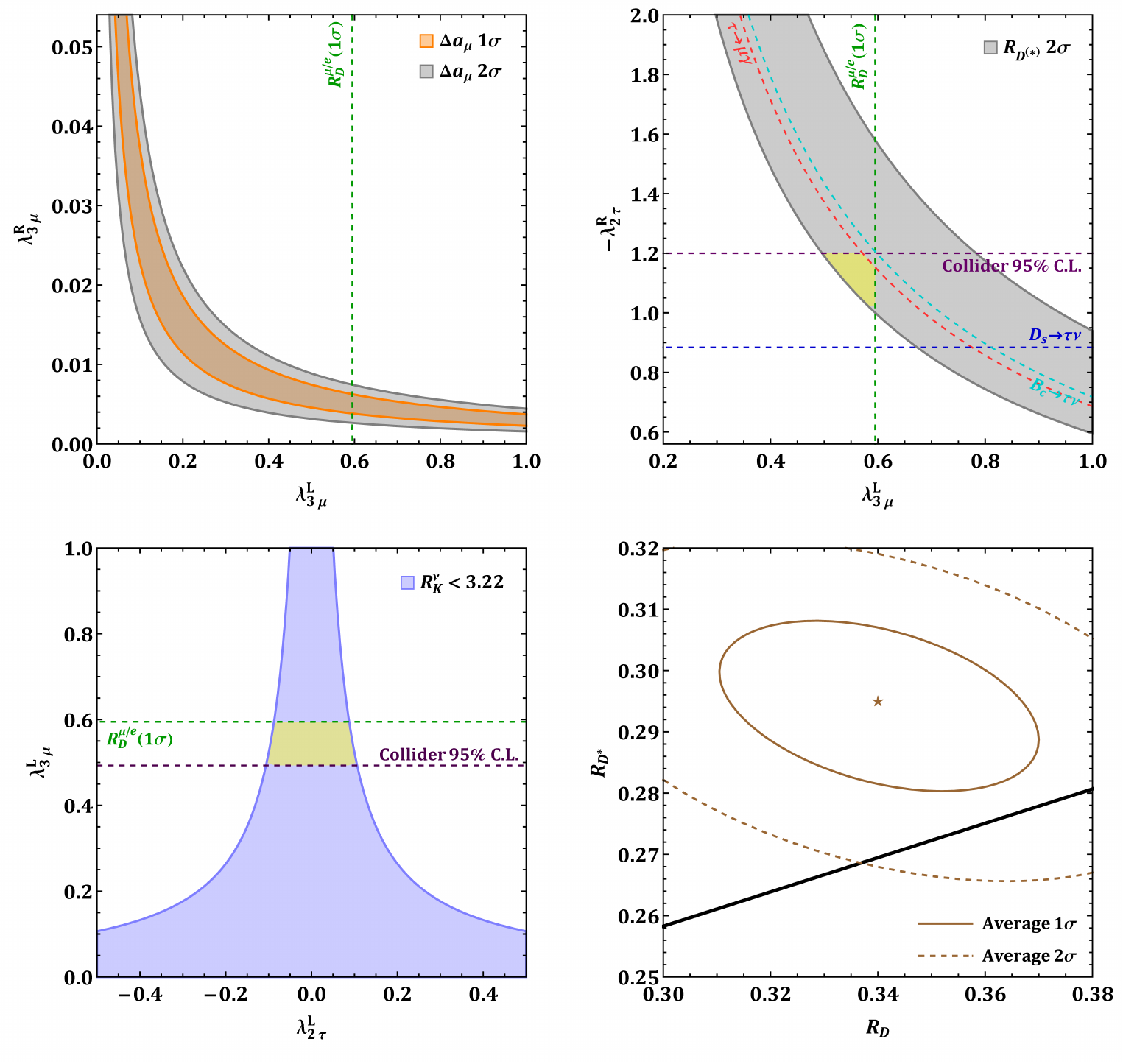} 	\vspace{-0.6cm}
	\caption{Allowed regions of the leptoquark Yukawa couplings consistent with the experimental measurements of $\Delta a^{}_\mu$ and $R^{}_{D^{(\ast)}_{}}$ at the $1\sigma$ (orange region) and $2\sigma$ (gray regions) levels are presented in the top two panels. The limits of other observables are labeled by dashed curves with different colors. In the bottom-left panel, the constraint from $B \rightarrow K^{(\ast)}_{}\nu\nu$ on the parameter space of $\{\lambda^{\rm L}_{2\tau},\lambda^{\rm L}_{3\mu}\}$ is represented by the blue-shaded area. The black curve in the bottom-right panel denotes the prediction for $R^{}_D$ and $R^{}_{D^{\ast}_{}}$ in our model, where the best-fit values (brown star), together with $1\sigma$ (solid ellipse) and $2\sigma$ (dashed ellipse) uncertainties of the averaged $R^{}_D$ and $R^{}_{D^{\ast}_{}}$ from Ref.~\cite{Amhis:2019ckw} are also exhibited for comparison. In addition, the yellow-shaded regions stand for the allowed parameter space obtained by combining all the observables together. Constraints from $\delta g^{Z}_{\mu{\rm L}}$ and $\delta g^{Z}_{\tau{\rm R}}$ are rather weak and not shown here.}
	\label{fig:LQ} 
	\vspace{0cm}
\end{figure}

As for the $Z^\prime_{}$ boson, the allowed parameter space of $m^{}_{Z^\prime_{}}/g^{}_{Z^\prime_{}}$ and $-\theta^{}_{\rm b}$ consistent with the observed $R^{}_{K^{(\ast)}_{}}$ anomaly are presented in the left panel of Fig.~\ref{fig:C9}, which indicates that $m^{}_{Z^\prime_{}}/g^{}_{Z^\prime_{}}$ becomes larger as $|\theta^{}_{\rm b}|$ increases. In Fig.~\ref{fig:C9} we also exhibit the constraints from meson mixing and $B \rightarrow K^{(*)}_{}\nu\nu$. The $B^{}_s-\overline{B}^{}_s$ mixing precludes large values of $m^{}_{Z^\prime_{}}/g^{}_{Z^\prime_{}}$ and $|\theta^{}_{\rm b}|$, leading to $m^{}_{Z^\prime_{}}/g^{}_{Z^\prime_{}} \lesssim 8~{\rm TeV}$ and $|\theta^{}_{\rm b}| \lesssim 0.16$ at the $1\sigma$ level, whereas the constraints from $D^{0}_{}-\overline{D}{}^0_{}$ mixing and $B \rightarrow K^{(*)}_{}\nu\nu$ are much weaker. 

Now let us focus on $S^{}_1$. The mass of $S^{}_1$ is taken as $m^{}_{S^{}_1} = 1~{\rm TeV}$, which can satisfy collider constraints and avoid non-perturbative leptoquark Yukawa couplings. We also set $\theta^{}_{\rm b}$ to be $-0.005$ in order to evade constraints from $\tau \rightarrow \mu\gamma$ and $B \rightarrow K^{(\ast)}_{}\nu\nu$. Then we search for the allowed parameter space of four leptoquark Yukawa couplings $\{\lambda^{\rm L}_{2\tau},\lambda^{\rm L}_{3\mu},\lambda^{\rm R}_{2\tau},\lambda^{\rm R}_{3\mu}\}$ by considering the constraints from $\Delta a^{}_\mu$, $R^{}_{K^{(\ast)}_{}}$, $R^{}_{D^{(\ast)}_{}}$ and other relevant observables. All the leptoquark Yukawa couplings are set to be real without loss of generality.

The results are displayed in Fig.~\ref{fig:LQ}. In the top-left panel, we demonstrate the correlation between $\lambda^{\rm L}_{3\mu}$ and  $\lambda^{\rm R}_{3\mu}$. $\lambda^{\rm R}_{3\mu}$ is inversely proportional to $\lambda^{\rm L}_{3\mu}$ owing to the dependence of $\Delta a^{}_\mu$ on these two parameters. Then the allowed parameter space of $\lambda^{ \rm L}_{3\mu}$ gives rise to small values of $\lambda^{\rm R}_{3\mu}$ with $ \lambda^{\rm R}_{3\mu} \simeq  {\cal O}(0.01)$. From the top-right panel of Fig.~\ref{fig:LQ}, we could see that the allowed ranges of $\lambda^{\rm L}_{3\mu}$ and $\lambda^{\rm R}_{2\tau}$ suffer strong constraints from the processes such as $\tau \rightarrow \mu \gamma$ and the direct leptoquark searches at LHC. In particular, if the $95\%$ C.L. upper limit on ${\cal B}(\tau \rightarrow \mu \gamma)$ is taken into account, the allowed parameter space of $\{\lambda^{\rm L}_{3\mu},\lambda^{\rm R}_{2\tau}\}$ will be compressed into a narrow belt. If we combine all the observables together, only the yellow region with $0.49 \lesssim \lambda^{\rm L}_{3\mu} \lesssim 0.59$ and $-1.20 \lesssim \lambda^{\rm R}_{2\tau} \lesssim -1.00$ survives and is still in agreement with the observed $R^{}_{D^{(\ast)}_{}}$ at the $2\sigma$ level. Moreover, we have checked that if $|\theta^{}_{\rm b}| \gtrsim 0.01$, 
there will be no allowed parameter space that can simultaneously explain the $R^{}_{D^{(\ast)}_{}}$ anomaly and accommodate other constraints. This observation in turn reduces the upper bound on $m^{}_{Z^\prime_{}}/g^{}_{Z^\prime_{}}$ to be $2~{\rm TeV}$, which implies the heavy $Z^\prime_{}$ gauge boson may be under strict constraints from the direct $Z^\prime_{}$ searches at the LHC. Assuming $\theta^{}_{\rm b} = -0.005$, in the right panel of Fig.~\ref{fig:C9} we plot the LHC constraints on the allowed parameters space of $m^{}_{Z^\prime_{}}$ and $g^{}_{Z^\prime_{}}$ by measuring the Drell-Yan process $q\overline{q} \to Z^\prime_{} \to \mu^+_{}\mu^{-}_{}$. The bound is taken from Fig.~9 of Ref.~\cite{Allanach:2021gmj} and rescaled to fit our model. We also show relation between $m^{}_{Z^\prime_{}}$ and $g^{}_{Z^\prime_{}}$ when $\Delta C^\mu_9$ takes its best-fit value in this plot. One can immediately find that the $Z^\prime_{}$ gauge boson with $m^{}_{Z^\prime_{}} \gtrsim 3.4~{\rm TeV}$ and $g^{}_{Z^\prime_{}} \gtrsim 2.6$ can simultaneously explain the $R^{}_{K^{(\ast)}_{}}$ anomaly and escape the constraints from direct $Z^\prime_{}$ searches at LHC. As the gauge coupling $g^{}_{Z^\prime_{}}$ is predicted to be relatively large, our model can be easily examined in the future collider experiments.

Furthermore, as can be seen from the bottom-left panel of Fig.~\ref{fig:LQ}, the experimental measurement of $R^\nu_{K^{(\ast)}_{}}$ constrains the parameter space of $\{\lambda^{\rm L}_{2\tau},\lambda^{\rm L}_{3\mu}\}$. Different from other Yukawa couplings, $\lambda^{\rm L}_{2\tau}$ is almost irrelevant to the explanations of muon $(g-2)$, $R^{}_{K^{(\ast)}_{}}$ and $R^{}_{D^{(\ast)}_{}}$ anomalies, and can be taken as zero without obeying the limits on other observables.

Finally, we also plot the correlation between $R^{}_D$ and $R^{}_{D^{\ast}_{}}$ in the bottom-right panel of Fig.~\ref{fig:LQ}, where we can find that the values of $R^{}_{D^{(\ast)}_{}}$ predicted in our model can only be compatible with the averaged results from Ref.~\cite{Amhis:2019ckw} at the $2\sigma$ level. Actually the correlation between $R^{}_D$ and $R^{}_{D^\ast_{}}$ is almost independent from the leptoquark Yukawa couplings. To make this point clearer, we neglect the terms proportional to $(\lambda^{\nu{\rm L}\ast}_{3\tau}\lambda^{\rm R}_{2\tau})/\widehat{m}^{2}_{S^{}_1}$ in the first two equations of Eq.~(\ref{eq:obratio}). Then we can eliminate $(\lambda^{\nu{\rm L}\ast}_{3\mu}\lambda^{\rm R}_{2\tau})^2_{}/\widehat{m}^{4}_{S^{}_1}$ by using these two equations, and arrive at $R^{}_{D^{\ast}_{}} \approx 0.280 R^{}_D + 0.174$. Hence it is impossible for us to adjust the values of free parameters to make both $R^{}_D$ and $R^{}_{D^\ast_{}}$ consistent with the experimental observations at the $1\sigma$ level.

\section{Summary}\label{sec:sum}
The latest muon $(g-2)$ result and the experimental measurements of $B$-meson decays reveal a series of flavor anomalies, which might be the indication of the LFU violation, and need to be confirmed by more experimental data. In order to simultaneously explain these anomalies, we construct a viable model by introducing an anomaly-free $U(1)^{}_X$ gauge symmetry with $X=B^{}_3 - 2L^{}_\mu/3 - L^{}_\tau/3$ together with an $S^{}_1$ leptoquark. Meanwhile, the spontaneous breaking of the $U(1)^{}_X$ symmetry by an additional SM singlet can generate realistic flavor mixing of quarks and leptons. Our main results are summarized as follows.

First, after the spontaneous breakdown of the $U(1)^{}_X$ and $SU(2)^{}_{\rm L} \otimes U(1)^{}_{\rm Y}$ symmetries, the up- and down-type quark matrices can be diagonalized via $V^{\rm L \dag}_{\rm u}M^{}_{\rm u}V^{\rm R}_{\rm u} = {\rm Diag}\{m^{}_u,m^{}_c,m^{}_t\}$ and $V^{\rm L \dag}_{\rm d}M^{}_{\rm d}V^{\rm R}_{\rm d} = {\rm Diag}\{m^{}_d,m^{}_s,m^{}_b\}$, where both $V^{\rm L}_{\rm u}$ and $V^{\rm L}_{\rm d}$ are the products of $(2,3)$- and $(1,2)$-rotations, and $V^{\rm R}_{\rm u,d}$ are the identity matrices. In the neutrino sector, the effective neutrino mass matrix $M^{}_\nu$ takes the ${\bf A}^{}_2$ two-zero texture, which predicts the normal mass ordering of three light neutrinos and relatively large CP violation. The $1\sigma$ allowed parameter space of free parameters is given.

Second, the heavy $Z^\prime_{}$ gauge boson associated with the $U(1)^{}_X$ symmetry mediates the $b \rightarrow s\mu^+_{}\mu^-_{}$ process at the tree level, and is used to account for the $R^{}_{K^{(\ast)}_{}}$ anomaly. The $S^{}_1$ leptoquark in our model is responsible for the explanations of muon $(g-2)$ and $R^{}_{D^{(\ast)}_{}}$ anomalies. Since the leptoquark coupling to the $b$-quark and $\nu^{}_\tau$ is forbidden by the gauge symmetry, the dominant process relevant for $R^{}_{D^{(\ast)}_{}}$ is $b \rightarrow c\tau\nu^{}_\mu$. Constraints from other processes mediated by $Z^\prime_{}$ or $S^{}_1$ are also discussed.

Third, we implement a detailed numerical analysis, and find that a $Z^\prime_{}$ boson with $m^{}_{Z^\prime_{}} \gtrsim 3.4~{\rm TeV}$ and $g^{}_{Z^\prime_{}} \gtrsim 2.6$ can successfully explain the $R^{}_{K^{(\ast)}_{}}$ anomaly and satisfy the restrictions from meson mixing and the direct $Z^\prime_{}$ searches at LHC. For the $S^{}_1$ leptoquark, taking all the relevant observables into consideration, we obtain the allowed parameter space of leptoquark Yukawa couplings  $\{\lambda^{\rm L}_{2\tau},\lambda^{\rm L}_{3\mu},\lambda^{\rm R}_{2\tau},\lambda^{\rm R}_{3\mu}\}$ that can accommodate the latest muon $(g-2)$ result at the $1\sigma$ level. The value of $R^{}_{D^{\ast}_{}}$ almost linearly depends on $R^{}_{D}$ with the relation $R^{}_{D^{\ast}_{}} \approx 0.280 R^{}_D + 0.174$, which implies the predictions for $R^{}_{D^{(\ast)}_{}}$ can only be compatible with the observed values at the $2\sigma$ level. Moreover, the LFV process $\tau \rightarrow \mu \gamma$ turns out to be the most stringent constraint in our model.

The $U(1)^{}_X$-gauged leptoquark models provide us with more space to address the flavor anomalies. Also, the gauge symmetry tightly constrains the leptoquark Yukawa couplings to fermions, and reduces free model parameters to a large degree. It is interesting to extend this study to other types of leptoquarks. We hope to come back to this issue in the near future.

\section*{Acknowledgements}
The author is greatly indebted to Prof.~Shun Zhou for inspiring comments and carefully reading this manuscript. The author also thanks Prof.~Xun-jue Xu, Bingrong Yu and Di Zhang for useful discussions. This work was supported in part by the National Natural Science Foundation of China under grant No. 11775232 and No. 11835013, by the Key Research Program of the Chinese Academy of Sciences under grant No. XDPB15, and by the CAS Center for Excellence in Particle Physics. All the Feynman diagrams in this work were produced by using JaxoDraw~\cite{Binosi:2003yf}.

\end{document}